\pdfoutput=1

\documentclass[11pt]{article}

\usepackage[preprint]{acl}

\usepackage{times}
\usepackage{latexsym}

\usepackage[T1]{fontenc}

\usepackage[utf8]{inputenc}

\usepackage{microtype}

\usepackage{inconsolata}

\usepackage{graphicx}

\usepackage{stmaryrd}
\usepackage{amsfonts}
\usepackage{amssymb}
\usepackage{amsmath}
\usepackage{mathtools}
\usepackage{bbm}
\usepackage{xspace}
\usepackage{todonotes}

\usepackage{algorithm}
\usepackage{algpseudocode}

\usepackage{tabularx,booktabs,multirow}  %
\usepackage{subcaption}
\usepackage{enumerate}
\usepackage{siunitx}

\usepackage{amsmath,amsfonts,bm}

\def\eqref#1{equation~\ref{#1}}
\def\Eqref#1{Equation~\ref{#1}}

\def\1{\bm{1}}

\def\rmT{{\mathbf{T}}}

\def\vc{{\bm{c}}}

\def\ve{{\bm{e}}}

\DeclareMathAlphabet{\mathsfit}{\encodingdefault}{\sfdefault}{m}{sl}
\SetMathAlphabet{\mathsfit}{bold}{\encodingdefault}{\sfdefault}{bx}{n}

\newcommand{\ourattack}{{paraphrasing attack}\xspace}
\newcommand{\ourdefence}{\textit{WET}\xspace}

\newcommand{\firstWM}{\text{EmbMarker}\xspace}
\newcommand{\secondWM}{\text{WARDEN}\xspace}

\newcommand{\fscore}{$F_1$-score\xspace}

\newcommand{\enron}{\text{Enron}\xspace}
\newcommand{\sst}{\text{SST2}\xspace}
\newcommand{\agnews}{\text{AG News}\xspace}
\newcommand{\mind}{\text{MIND}\xspace}

\newcommand{\gpt}{\text{GPT-3.5}\xspace}
\newcommand{\nllb}{\text{NLLB}\xspace}
\newcommand{\dipper}{\text{DIPPER}\xspace}

\newcommand{\refalg}[1]{Algorithm~\ref{#1}}
\newcommand{\refthm}[1]{Theorem~\ref{#1}}
\newcommand{\refapp}[1]{Appendix~\ref{#1}}

\newcommand{\reffig}[1]{Figure~\ref{#1}}
\newcommand{\refsec}[1]{Section~\ref{#1}}
\newcommand{\reftab}[1]{Table~\ref{#1}}

\usepackage{ntheorem}

\newtheorem{theorem}{Theorem}

\theoremstyle{nonumberplain}
\newtheorem{proof}{Proof}

\def\eg{{\em e.g.,}\xspace}
\def\ie{{\em i.e.,}\xspace}

\def\aka{{\em a.k.a}\xspace}

\definecolor{OliveGreen}{rgb}{0,0.6,0}

\title{WET: Overcoming Paraphrasing Vulnerabilities in Embeddings-as-a-Service with Linear Transformation Watermark}

\author{
  Anudeex Shetty$^{1}$,  Qiongkai Xu$^{1,2}$, Jey Han Lau$^{1}$ \\
  $^1${School of Computing and Information System, the University of Melbourne, Australia} \\
  $^2${School of Computing, FSE, Macquarie University, Australia} \\
  {\tt anudeexs@student.unimelb.edu.au } \\
  {\tt qiongkai.xu@mq.edu.au } \\
  {\tt laujh@unimelb.edu.au }
}

\begin{document}
\maketitle
\begin{abstract}
Embeddings-as-a-Service (EaaS) is a service offered by large language model (LLM) developers to supply embeddings generated by LLMs. Previous research suggests that EaaS is prone to imitation attacks---attacks that clone the underlying EaaS model by training another model on the queried embeddings. As a result, EaaS watermarks are introduced to protect the intellectual property of EaaS providers. In this paper, we first show that existing EaaS watermarks can be removed by paraphrasing when attackers clone the model. Subsequently, we propose a novel watermarking technique that involves linearly transforming the embeddings, and show that it is empirically and theoretically robust against paraphrasing.\footnote{The code can be found at \url{https://github.com/anudeex/WET.git}.}
\end{abstract}

\section{Introduction}
Large language models (LLMs) represent the state-of-the-art in natural language processing (NLP) due to their remarkable ability to understand languages and generate texts~\cite{zhao2023survey-LLM}. To make LLMs more accessible, LLM developers such as OpenAI and Google provide Machine-Learning-as-a-Service (MLaaS) to assess their models.

Embeddings-as-a-Service (EaaS) is a variant of MLaaS that offers feature extraction capabilities by delivering embeddings generated by LLMs \citep{2022-EaaS-OpenAI}. 
Alarmingly, \citet{stolen-encoder} demonstrated successful imitation attacks on these services. Specifically, they showed that it is possible to clone the underlying EaaS model by training a different model (with a different architecture) using queried embeddings, thereby violating the intellectual property (IP) of LLM developers.

Watermarking techniques have been proposed to defend against these EaaS imitation attacks. %
EmbMarker \citep{peng2023you} introduces a method that integrates a \textit{target embedding} into the original embedding based on the presence of \textit{trigger words}---a pre-defined set of words---in the input text. Such techniques implant verifiable statistical signals, \ie watermarks, for the service provider to verify if their model has been copied. 
However, \citet{shetty2024warden} demonstrated that an attacker could circumvent EmbMarker by using
a contrastive method to identify and remove the single target embedding from the embedding space. To counter this, they introduced \secondWM, which strengthens the defence by incorporating \textit{multiple} target embeddings instead of just one, making it more challenging for an attacker to eliminate the watermarks.

Nonetheless, these methods rely on words to trigger watermark injection, which we suspect could be circumvented by paraphrasing the input texts and using their queried embeddings during imitation attacks. To this end, we show that paraphrasing does dilute the watermark and thereby reveals a new form of vulnerability in these watermarking techniques. 
To address this vulnerability, we introduce a new defence technique, \ourdefence (\textbf{W}atermarking \textbf{E}aaS with Linear \textbf{T}ransformation), which applies linear transformations to the original embeddings to implant watermarks that can be verified later through reverse transformation. We analyse \ourdefence both theoretically and empirically to show it is robust against the new paraphrasing attack. Extensive experiments demonstrate near-perfect verifiability, even with one sample.
Additionally, the utility of embeddings is mostly preserved due to the use of simple linear transformations.

The contributions of our work are as follows:
\begin{itemize}
    \item We introduce and validate {paraphrasing} attack to bypass current EaaS watermarking techniques.
    \item We design a novel EaaS watermarking method, \ourdefence, and show that it is robust against paraphrasing attacks. %
\end{itemize}

\section{Related Work}

\subsection{Imitation Attacks}
An imitation attack, also known as ``model stealing'' or ``model extraction'' \citep{tramer2016stealing,orekondy2019knockoff,krishna2019thieves,wallace-etal-2020-imitation}, involves an imitator querying an MLaaS (or EaaS) to construct a surrogate model without the authorisation of the victim service providers. The primary motivation is to bypass service charges or even offer competitive services \citep{xu-he-2023-security}. Imitation attacks extend beyond IP violations; they can also be used to craft adversarial examples \citep{he-etal-2021-model} and conduct privacy breaches like attribute inference \citep{he-etal-2022-extracted}. Notably, \citet{xu-etal-2022-student} demonstrated that a copied model can outperform the victim model through ensemble and domain adaptation. Recent successful imitation attacks on EaaS \citep{stolen-encoder} not only compromise the confidentiality of embeddings but also violate the copyright of EaaS providers. These attacks constitute the threat model we explore in our research.

\subsection{Text Watermarks}
\citet{he2022protecting} and \citet{he2022cater} introduced early text watermarking techniques by selectively replacing words in LLM-generated text with synonyms. A more recent work by \citet{kirchenbauer2023watermark} advanced text watermarks by biasing LLMs towards a set of preferred words---verifiable later---using a pseudo-random list based on the most recent tokens. Building on this approach, several works \citep{kuditipudi2023robust, christ2023undetectable, openAI-LLM-WM} have applied cryptographic methods to watermarking, using a secret key to minimise the gap between original and watermarked distributions, thereby making the watermark unbiased and stealthy.

Recent studies \citep{sadasivan2023can,krishna2024paraphrasing} demonstrated that these watermarks are vulnerable to paraphrasing-based attacks, where paraphrasing the generated text disrupts the token sequences, thereby evading watermark detection. Similarly, \citet{RTT-wm} demonstrated that round-trip translation, another form of paraphrasing, can diminish watermark detection. These observations motivate our attack, in which we explore the use of paraphrasing and round-trip translation to remove watermarks from \textit{embeddings}.

\subsection{Embedding Watermarks}
\label{sec:eaas-wms}
\citet{peng2023you} proposed EmbMarker, the first watermark algorithm designed to protect EaaS against imitation attacks. This algorithm uses a set of trigger words and a fixed target embedding as a watermark, where the target embedding is proportionally added to the original embedding based on the number of trigger words present in the input text. In other words, the number of trigger words determines the \textit{watermark weight}. 
However, EmbMarker has only been tested against a narrow range of attacks and relies on the secrecy of the target embedding.

 \citet{shetty2024warden} showed that it is possible to recover the target embedding used in EmbMarker and subsequently eliminate it from the embeddings. To counter this, they proposed WARDEN, an improved watermarking technique that incorporates multiple target embeddings, making the watermark more difficult to recover. WARDEN, however, still relies on trigger words and, therefore, might remain susceptible to paraphrasing during imitation attacks.

\section{Methodology}

We first provide an overview of the existing EaaS watermark techniques and their benefits in defending against imitation attacks. We then introduce our \ourattack and subsequently propose
a new watermarking technique, \ourdefence. %

\subsection{Preliminary Background}
\label{sec:preliminary}

\begin{figure*}[!th]
    \centering
    \includegraphics[width=0.99\textwidth,keepaspectratio]{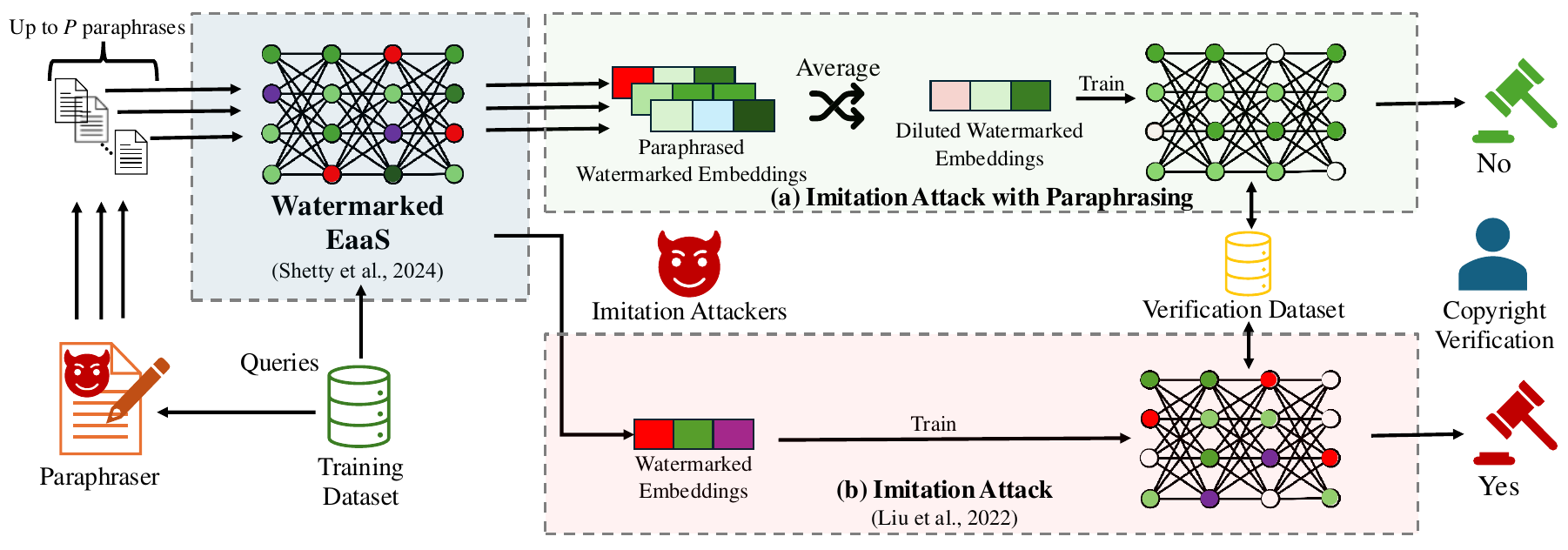}
    \caption{An overview of our paraphrasing attack, where the (a) \textcolor{OliveGreen}{Green} area shows the EaaS watermarks (presented as the elements in \textcolor{red}{Red}) getting diluted due to paraphrasing and potentially bypassed. On the contrary, the (b) \textcolor{red}{Red} area denotes a traditional imitation attack without paraphrasing, leading to copyright infringement. 
    }
    \label{fig:attack-overview}
\end{figure*}

We assume that a malicious attacker conducts an imitation attack on a victim EaaS service $\mathbb{S}_v$ based on model $\Theta_v$. The attacker queries $\mathbb{S}_v$ to collect the embeddings (which are watermarked, unbeknownst to the attacker) for a set of input texts $D_a$, which will then be used for training an attack/surrogate model $\Theta_a$. The goal of the attacker is to provide a competitive EaaS service $\mathbb{S}_a$ and they may actively employ strategies to remove or bypass the watermark. 
For the victim, \ie EaaS provider, it is crucial that the watermarked embeddings perform similarly to the original non-watermarked embeddings on downstream tasks. To determine whether their model has been copied, the victim will query suspicious services $\mathbb{S}_a$ to check if the returned embeddings contain the injected watermarks.

\subsection{Paraphrasing Attack}

We propose generating multiple paraphrases and using their averaged embedding to train the surrogate model so as to bypass the detection of embedding watermark; see \reffig{fig:attack-overview} for an illustration.

Formally, we generate $P$ paraphrased texts 
$S_P = \{ s^{1},\cdots,s^{P} \}$ 
given an {input text} $s$. 
Next, we query the EaaS $\mathbb{S}_{v}$ to get their embeddings and aggregate them into a single embedding through averaging ($\text{avg}(\cdot)$):
    \begin{align*}
        E_a &= \bigl\{ \mathbb{S}_{v}(s^{ i }) \bigr\}_{i=1}^{P}, \ 
        \text{avg}(E_a) = \sum\limits_{\ve \in E_a}\ve / {|E_a|}.
    \end{align*}

We will then use the aggregated embeddings $\text{avg}(E_a)$ for training the surrogate model in an imitation attack (illustrated as the ``Diluted Watermarked Embeddings'' in \reffig{fig:attack-overview}). 
To measure the success of this paraphrasing attack, we will evaluate verification accuracy (\textit{verifiability}) and downstream task performance (\textit{utility}), as detailed in \refsec{sec:metrics}. The attack would be considered successful if the downstream task performance is high and verification accuracy is low.

We provide a theoretical validation in the \refapp{app:attack-theoretical-proof} to show that averaging paraphrase embedding reduces the possibility of observing embedding samples with high watermark weights. We validate this hypothesis empirically in \refsec{sec:empirical-backdoor-analysis}.

\begin{algorithm}[t]
\caption{Transformation Matrix Generation.}
\begin{algorithmic}[1]
\Require 
\Statex n: \small \# original dimensions \normalsize
\Statex k: \small \# original dimensions used in transformation \normalsize
\Statex w: \small \# watermarked embedding dimensions \normalsize
\Function{Matrix\_Gen}{$n,k$}
    \State Initialise $\rmT \gets \phi$
    
    \State $\text{row} \gets \Call{Row\_Gen}{n,k}$ \Comment{$\mathbb{R}^{1 \times n}$}
    \State $\text{cnt} \gets 0$
    \For{each $i = 1, 2, \cdots, w$} \Comment{\small Circular \normalsize}
        \State $\rmT[i] \gets \text{row}$
        \State $\text{row} \gets \text{Roll}(\text{row})$ 
        \State $\text{cnt } \mathrel{+}= 1$
        \If{$\text{cnt} == n$} \Comment{\small Re-generate \normalsize}
            \State $\text{row} \gets \Call{Row\_Gen}{n,k}$ 
            \State $\text{cnt} \gets 0$
        \EndIf
    \EndFor
    
    \State \Return $\rmT$ \Comment{$\mathbb{R}^{w \times n}$}
\EndFunction

\item[]
\Function{Row\_Gen}{$n,k$}
    \State Initialise $\text{row} \gets \text{Zeroes}(n)$
    \State $\text{positions} \gets \text{Sample}(n,k)$  \Comment{\small Correlations \normalsize}
    
        \For{$\text{p}$ in $ \text{positions}$}
            \State $\text{row}[\text{p}] \sim U(0,1)$ \Comment{\small Random \normalsize}
        \EndFor 
        \State $\text{row} \gets \text{Norm}(\text{row})$
        \State \Return $\text{row}$
\EndFunction
\end{algorithmic}
\label{algo:matrix-generation}
\end{algorithm}

\subsection{\ourdefence Defence}
\label{sec:wet-defense}

\begin{figure*}[t]
    \centering
    \includegraphics[width=\textwidth,keepaspectratio]{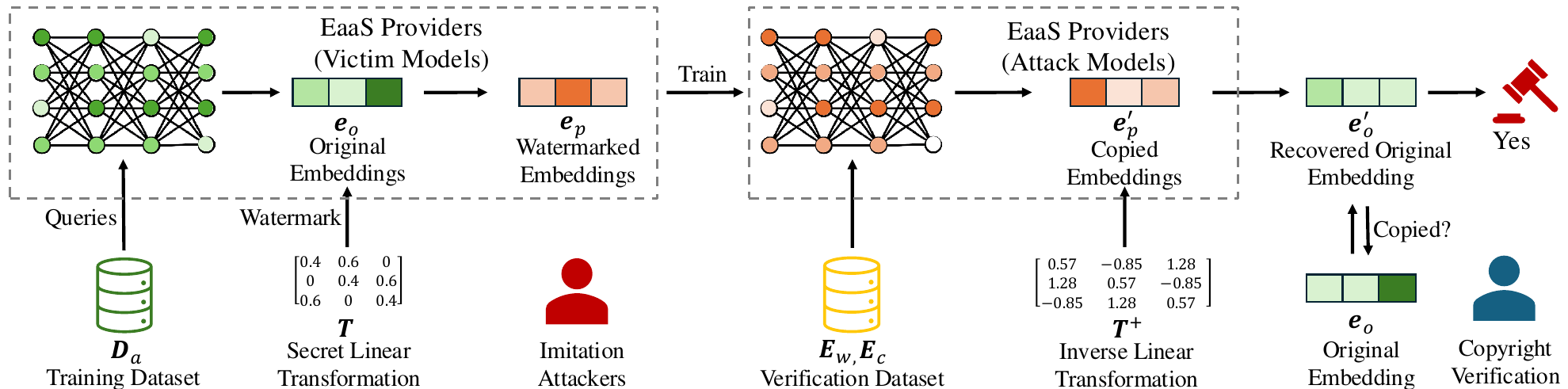}
    \caption{An overview of the workflow for \ourdefence. The left block illustrates the watermarking process using a secret transformation matrix $\rmT$. The right block follows the watermark verification process, employing the pseudoinverse of the transformation matrix $\rmT^+$. The recovered embedding $\ve_{o}^{\prime}$ and the original embedding $\ve_o$ are compared in copyright verification.}
    \label{fig:defence-overview}
\end{figure*}

Next, we introduce \textbf{W}atermarking \textbf{E}aaS with Linear \textbf{T}ransformation (\ourdefence), a new embedding watermarking protocol (shown in \reffig{fig:defence-overview}) that is designed to be robust against paraphrasing attacks. The core idea is to use a preset linear transformation matrix  $\rmT$ (unknown to the attacker) to transform an original embedding $\ve_o$ into a watermarked embedding $\ve_p$ (\reffig{fig:defence-overview} left part). Our watermarking technique discards the original elements and retains only the transformed ones, which makes the watermark more difficult to be detected.\footnote{We initially explored embedding dimension obfuscation by adding new dimensions mixed with the original ones, inspired by \citet{yan2023rethinking}, but found that these obfuscated dimensions could be easily identified using feature correlation and feature importance techniques; details in \refapp{sec:hyp-obfuscation}.} To check for the watermark in a copied embedding $\ve'_p$ (produced by the surrogate model), we apply the \textit{inverse} of the linear transformation matrix $\rmT^{+}$ to it and assess whether the recovered embedding $\ve'_{o}$ is similar to the original embedding $\ve_{o}$ (\reffig{fig:defence-overview} right part). 
An important consideration is constructing the transformation matrix in a way that balances the trade-off between utility and verifiability.

\paragraph{Watermark Injection.}

Given a transformation matrix $\rmT$, we (i) multiply it with the original embedding $\ve_o$ and (ii) normalise it to a unit vector, %
\begin{equation}
    \centering
    \begin{aligned}
        \ve_{p} = \text{Norm}(\rmT \cdot \ve_{o}) = \frac{\rmT \cdot \ve_{o}}{|| \rmT \cdot \ve_{o} ||} .
    \end{aligned}
    \label{eq:wm-injection}
\end{equation}

Note that, unlike previous EaaS watermarks, our approach does not rely on trigger words for watermark injection. Instead, we watermark all the output embeddings, leading to denser signals and making it more difficult to bypass while maintaining the same level of utility for EaaS users.

\paragraph{Matrix Construction.}
\label{sec:matrix-construction}

One challenge for \ourdefence is in designing the transformation matrix. In the watermark verification process, we perform a reverse transformation (\Eqref{eq:reverse-transformation}) to recover the original embeddings from the watermarked ones. There, it is crucial that the transformation matrix is both full-rank and well-conditioned to allow for accurate pseudoinverse computation \citep{strang2000linear}. To meet the requirement, we adopt circulant matrices \citep{circulant-matrix} to ensure these properties. The first row is generated randomly, and subsequent rows are circulations of the initial row. The positions and values of non-zero entries in the first row are selected randomly (see ``Secret Linear Transformation'' matrix in \reffig{fig:defence-overview}). 
The circulant matrix is full-rank if the first row has non-zero fast Fourier transform (FFT) values (corresponding to eigenvalues of circulant matrix), which is more probable by our row construction \citep{circulant-matrix}. Moreover, full-rank guarantees a lower condition number, which is beneficial for computing well-conditioned pseudoinverses \citep{strang2000linear}. Additionally, cycle shifts ensure that 
all dimensions in the original embedding contribute equally to the watermark. \refalg{algo:matrix-generation} details the generation of the transformation matrix.
The two hyperparameters to be considered are $w$ and $k$. $w$ represents the number of dimensions of the watermarked embeddings. $k$ represents the number of original dimensions used to compute a dimension in the watermarked embeddings. 
We explore and discuss alternative matrix constructions by relaxing various properties (like circularity, randomness, and others) in \refapp{app:diff-matrices}.

\paragraph{Robustness to Paraphrasing Attacks.}

We now show theoretically how the linear transformation used in \ourdefence is robust against paraphrasing during imitation attacks and the watermark is still learned by the surrogate model. %

\begin{theorem}[Robustness of \ourdefence]

Given $P$ watermarked embeddings, $\ve_{p}^{i} = f(\ve_{o}^{i})$, where $f$ is a linear transformation function, as defined in \Eqref{eq:wm-injection} and $i \in [1 .. P]$. The average of these paraphrased embeddings is equivalent to a linear transformation of a pseudo-aggregation of the original embeddings, $\hat{\ve}_{o}^{i}$, \ie
\begin{equation}
    \text{avg} ( f(\{\ve_{p}^{i} \}_{i=1}^{P}) ) = f(\text{avg} (\{\hat{\ve}_{o}^{i} \}_{i=1}^{P}).
\end{equation}

        \begin{proof}
            \begin{align*}
                \text{avg} ( f(\{\ve_{p}^{i} \}_{i=1}^{P}) ) = \text{avg} (\{{\underbrace{\text{Norm}(\rmT \cdot \ve_{o}^{i})}_{
                    \triangleq \alpha_{i} \cdot \rmT \cdot \ve_{o}^{i}}
                    }\}) \\
                = \rmT \cdot \frac{1}{P} \sum_{i=1}^{P} {\underbrace{\alpha_i \cdot \ve_{o}^{i}}_{\triangleq \hat{\ve}_{o}^{i}}}
                = \rmT \cdot \text{avg} (\{\hat{\ve}_{o}^{i} \}_{i=1}^{P}) \\
                = f(\text{avg} (\{\hat{\ve}_{o}^{i} \}_{i=1}^{P}).
            \end{align*}
    \end{proof}
\label{theorem:RQ2-proof}
\end{theorem}

The transformation $\rmT$ should be consistent regarding the aggregation on the pseudo embedding $\hat{\ve}$ though distorted by $\alpha_i=1/|| \rmT \cdot \ve_{o}^i ||$. Given \refthm{theorem:RQ2-proof}, the \ourdefence watermark key (\ie $\rmT$) will not be removed through the aggregation of paraphrased embeddings.

\paragraph{Watermark Verification.}
\label{sec:defence-verification}
The verification process attempts to decode the watermarked embedding using the authentic $\rmT$ and verify whether it matches the original embedding.
That is, we first apply the pseudoinverse of the transformation matrix $\rmT^{+}$ to the copied embedding $\ve'_{p}$ to produce recovered original embedding $\ve'_{o}$:
\begin{equation}
    \centering
    \begin{aligned}
        \ve'_{o} &= \rmT^{+} \cdot \ve'_{p},
    \end{aligned}
    \label{eq:reverse-transformation}
\end{equation}
where $\rmT^{+}$ is Moore-Penrose inverse (\aka pseudoinverse) \citep{strang2000linear}. When $\rmT$ has linearly independent rows (guaranteed by the circulant matrix), then $\rmT^{+}$ is a right inverse, \ie $\rmT \cdot \rmT^{+} = \mathbf I_w.$ %

To check the transformation aligns with the authentic watermark process, we measure the similarity between the recovered embedding $\ve'_{o}$ by the attack model and the original embedding $\ve_{o}$ by the victim model. If the attacker has copied the victim model, then the similarity score should be high. In our experiments, we use cosine similarity for measuring similarity.\footnote{Although $l_2$ distance is also used as a similarity metric conventionally, we found similar performances in our experiments and have omitted its results for brevity.}

\section{Experiments}
\begin{figure*}[h]
    \centering
    \begin{subfigure}{0.47\textwidth}
    \centering\includegraphics[width=\linewidth,keepaspectratio]{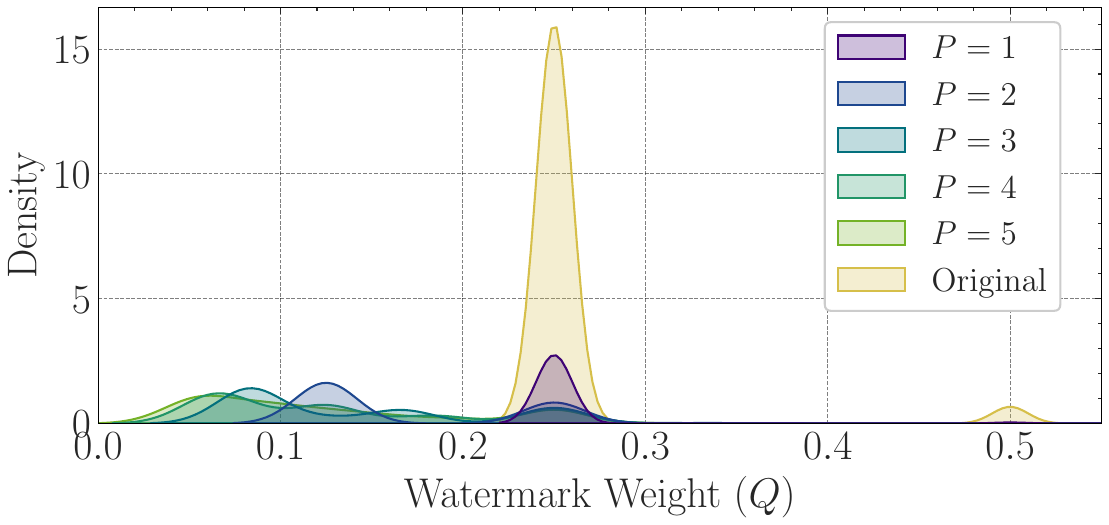}
    \caption{\enron}
    \end{subfigure}
    \begin{subfigure}{0.48\textwidth}
    \centering\includegraphics[width=\linewidth,keepaspectratio]{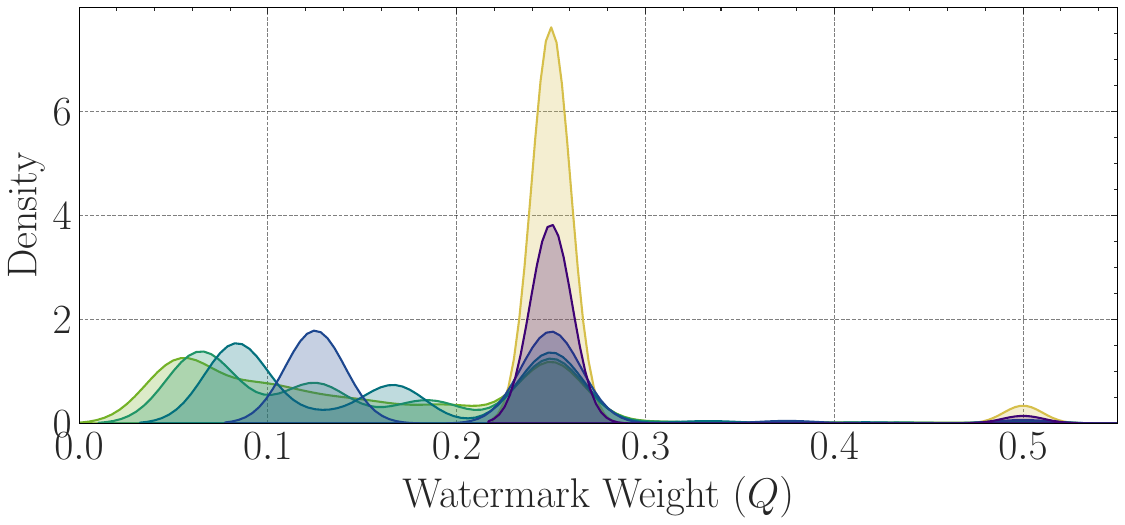}
    \caption{\sst}
    \end{subfigure}
    \begin{subfigure}{0.48\textwidth}
    \centering\includegraphics[width=\linewidth,keepaspectratio]{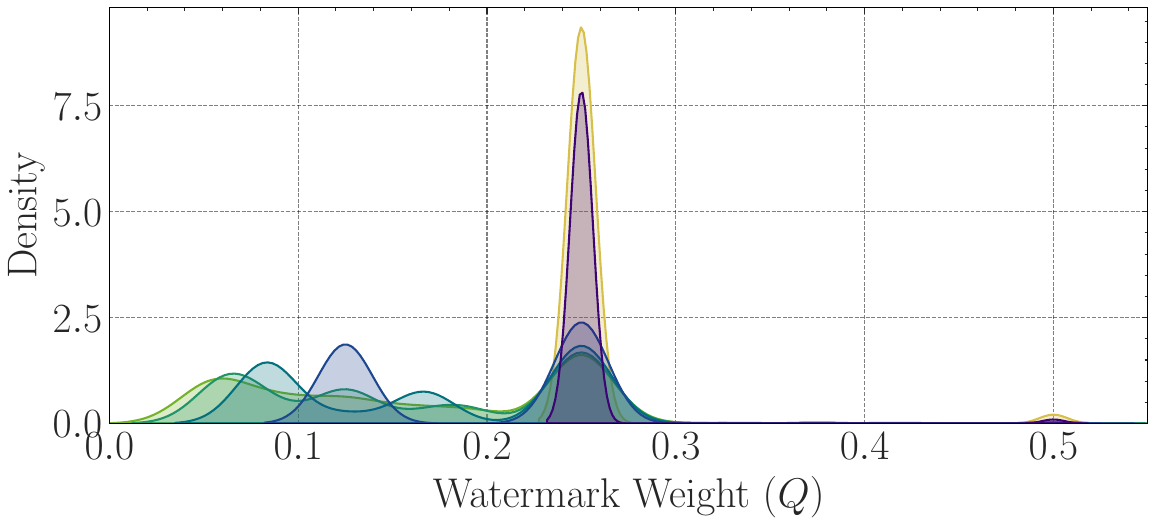}
    \caption{\mind}
    \end{subfigure}
    \begin{subfigure}{0.47\textwidth}
    \centering\includegraphics[width=\linewidth,keepaspectratio]{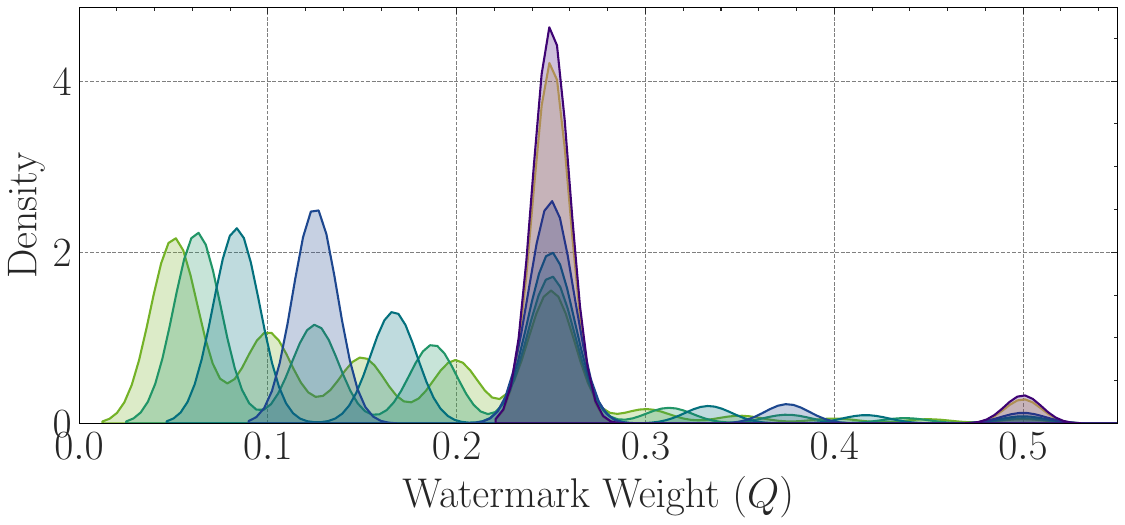}
    \caption{\agnews}
    \end{subfigure}
    \caption{Watermark weight analysis for different datasets (in subcaption) based on \gpt paraphrases. In general, aggregating watermarked embeddings from more paraphrases (larger $P$) reduces the watermark weights.}
    \label{fig:backdoor-level-analysis}
\end{figure*}

\subsection{Metrics}
\label{sec:metrics}
To evaluate the effectiveness of the paraphrasing attack and our new watermarking method, we use the following metrics to assess downstream task utility and watermark verifiability.

\paragraph{Downstream Task Utility.}

Using the EaaS embeddings as input, we build multi-layer perceptron classifiers for a range of classification tasks and evaluate the accuracy (ACC) and \fscore (F1) performance. This evaluation serves as an indicator of whether watermarking degrades the quality of the original embeddings: ideally, there should be minimal performance difference between the watermarked and original embeddings.

\paragraph{Watermark Verifiability.}

To quantify verification performance, we create a verification dataset containing two sets of embeddings: (i) watermark set $E_w$ (which contains watermarked embeddings) and (ii) contrast set $E_c$ (which contains watermarked embeddings generated with a different transformation matrix).\footnote{Note that for \firstWM and \secondWM, we follow the original studies where the contrast set are non-watermarked embeddings (\ie embeddings where their input text have no trigger words).}
The goal is that the verification process should have a high accuracy in identifying $E_w$ without confusing it with $E_c$.

Given the two sets, we compute the average cosine similarity between the recovered embeddings ($\ve_{o}^{\prime i }$ from \Eqref{eq:reverse-transformation}) and original embeddings ($\ve_{o}^{ i }$) and then take their difference:
\begin{equation}
    \centering
    \begin{aligned}
\Delta_{cos} = cos_\text{avg} (S_w) - cos_\text{avg} (S_c), \\
        cos_\text{avg} (S) = \frac{1}{|S|} \sum_{i=1}^{|S|} cos(\ve_{o}^{\prime i }, \ve_{o}^{ i }), \\
    \end{aligned}
    \label{eq:avg-cos-l2}
\end{equation}
where the sets of recovered and original embedding pairs are constructed by:
\begin{equation}
    \begin{aligned}
        S_w &= \left\{ ( \ve_{o}^{\prime i}, \ve_{o}^{ i} ) | \ve_{p}^{\prime i} \in E_{w} \right\}_{i=1}^{|E_w|}, \\
        S_c &= \left\{ ( \vc_{o}^{\prime i}, \vc_{o}^{i} ) | \vc_{p}^{\prime i} \in E_{c} \right\}_{i=1}^{|E_c|}.
    \end{aligned}
\end{equation}

Based on the cosine similarity scores, we also compute the area under the receiver operating characteristic curve (AUC) \citep{mitchell2023detectgpt}, which gives us a more intuitive interpretation of verifiability: an AUC of 100\% means the watermark set and contrast set are perfectly separable.
Additional details regarding the evaluation dataset are provided in \refapp{sec:pos-neg-samples}.

\subsection{Datasets}
\label{sec:datasets}
We use \agnews \citep{ag_news}, \mind \citep{mind}, \sst \citep{sst2}, and \enron \citep{enron} in our experiments. \reftab{table:dataset-statistics} provides the statistics for these datasets. These datasets are used to evaluate a variety of downstream classification performances, covering tasks from spam classification (\enron) to sentiment classification (\sst) to news recommendation and classification (\agnews and \mind).
\begin{table}[ht]
\small
    \centering
    \begin{tabular}{cccccc}

        \toprule
        {Dataset} & {\# Class} & {\# Train} & {\# Test} & {Avg. Len.} \\
        
        \midrule
        {\enron} & {2} & {31,716} & {2,000} & {34.57} \\
        {\sst} & {2} & {67,349} & {872} & {54.17} \\
        {\mind} & {18} & {97,791} & {32,592} & {66.14} \\
        {\agnews} & {4} & {120,000} & {7,600} & {236.41} \\
        \bottomrule
    \end{tabular}
    \caption{The statistics of datasets.}
    \label{table:dataset-statistics}
\end{table}

\subsection{Experimental Settings}
In terms of model configurations and hyper-parameters, we largely follow the experimental settings by \citet{shetty2024warden}. To simulate the imitation attacks, we use {GPT-3 text-embedding-002} \citep{2022-EaaS-OpenAI} as the victim EaaS to retrieve the original (non-watermarked) embeddings and BERT \citep{devlin-etal-2019-bert} as backbone model for the attacker's surrogate model.\footnote{All watermarking techniques (\firstWM, \secondWM and \ourdefence) inject watermarks post-hoc into the embeddings produced by the API calls.}
We experiment with three paraphrase methods: (i) prompting GPT-3.5-turbo (prompts are given in \refapp{app:gpt-exp-setting});
(ii) using an specialised paraphrasing model, DIPPER \cite{krishna2024paraphrasing} (configuration is detailed in \refapp{app:dipper-exp-setting}); and (iii) round-trip translation using NLLB \citep{NLLB}, a multi-lingual translation model. We present more details for these models (\eg pivot languages and translation setups) in \refapp{app:nllb-exp-setting}.
When paraphrasing, we first generate $P$ unique paraphrases for each input text and then filter out bad paraphrases based on their cosine similarity with the original input text (details in the \refapp{appendix:cos-sim-filter}). %
Consequently, on average, we have 2.68, 3.30, 3.41, and 4.89 paraphrases (using \gpt) across \enron, \sst, \mind, and \agnews, respectively.
\refapp{app:para-quality} presents more analyses on the quality of the paraphrases.
For our main experiments, we set $w=n$ (recall that $n$ and $w$ are the number of dimensions in the original and watermarked embeddings) to avoid compressing the embeddings. We investigate different values of $w$ in \refsec{sec:diff-h-defence}.

\subsection{Attack Experiments}
\label{sec:attack-exps}

We now present the results of our paraphrasing attack against \firstWM and \secondWM.

\paragraph{Watermark Weight Analysis.}
\label{sec:empirical-backdoor-analysis}

\reffig{fig:backdoor-level-analysis} shows the watermark weight distribution with varying numbers of paraphrases $P$ for \firstWM and \secondWM.\footnote{\firstWM and \secondWM use the same watermark weight so these results apply to both methods.} %
As $P$ increases, we observe that the watermark weight reduces, suggesting that the more paraphrases incorporated, the more diluted the watermark. In other words, these results paraphrasing might be able to circumvent the watermark detection for an imitation attack. 
We present other attack setups in Appendix~Figures~\ref{fig:nllb-wm-wt-analysis}~and~\ref{fig:dipper-wm-wt-analysis}.

\paragraph{Utility and Verifiability Evaluation.}
\label{sec:attack-perf}
\begin{table}[!t]
\centering
    \begin{minipage}{\columnwidth}
    \resizebox{\columnwidth}{!}{%
    \begin{tabular}
    {ccccc}
    \toprule
    {\textbf{Method}} & \textbf{ACC $\uparrow$}  & \textbf{F1 $\uparrow$}  & {\textbf{$\Delta_{cos} \downarrow$}} & {\textbf{AUC} $\downarrow$} \\
    \toprule

    {WARDEN} & 94.50$\pm$0.34 & 94.50$\pm$0.34 & 5.20$\pm$0.34 & 97.40$\pm$0.54 \\
    \midrule
    {+\gpt Attack} & 92.81$\pm$0.21 & 92.81$\pm$0.21 & 0.70$\pm$0.22 & 68.90$\pm$7.79 \\
    {+\dipper Attack} & 91.34$\pm$0.52 & 91.33$\pm$0.52 & 0.46$\pm$0.11 & 67.50$\pm$5.56 \\
    {+\nllb Attack} & 93.35$\pm$0.23 & 93.35$\pm$0.23 & 0.65$\pm$0.12 & 71.95$\pm$4.04 \\
    
    \bottomrule
    
    \end{tabular}}
    \vspace{-0.9em}
    \caption*{(a) {\enron}}
    \vspace{-0.1em}
    \end{minipage}

    \begin{minipage}{\columnwidth}
    \resizebox{\columnwidth}{!}{%
    \begin{tabular}
    {ccccc}
    \toprule

    {WARDEN} & 93.10$\pm$0.12 & 93.10$\pm$0.12 & 2.57$\pm$1.19 & 86.75$\pm$6.20 \\
    \midrule
    {+\gpt Attack} & 92.75$\pm$0.15 & 92.75$\pm$0.15 & 0.93$\pm$0.09 & 75.90$\pm$2.91 \\
    {+\dipper Attack} & 91.70$\pm$0.27 & 91.66$\pm$0.27 & 0.90$\pm$0.17 & 71.95$\pm$2.69 \\
    {+\nllb Attack} & 92.57$\pm$0.09 & 92.55$\pm$0.08 & 1.06$\pm$0.19 & 69.35$\pm$2.94 \\

    \bottomrule
    \end{tabular}}
    \vspace{-0.9em}
    \caption*{(b) {\sst}}
    \vspace{-0.1em}
    \end{minipage}
    
    \begin{minipage}{0.99\columnwidth}
    \resizebox{\columnwidth}{!}{%
    \begin{tabular}
    {ccccc}
    \toprule

    {WARDEN} & 77.31$\pm$0.08 & 51.47$\pm$0.23 & 5.27$\pm$0.17 & 98.10$\pm$0.51 \\
    \midrule
    {+\gpt Attack} & 77.01$\pm$0.05 & 51.24$\pm$0.22 & 1.85$\pm$0.21 & 79.40$\pm$3.08 \\
    {+\dipper Attack} & 76.86$\pm$0.07 & 50.54$\pm$0.17 & 3.47$\pm$0.12 & 96.70$\pm$0.51 \\
    {+\nllb Attack} & 76.64$\pm$0.10 & 50.36$\pm$0.11 & 3.89$\pm$0.06 & 97.80$\pm$0.33 \\

    \bottomrule
    \end{tabular}}
    \vspace{-0.9em}
    \caption*{(c) {\mind}}
    \vspace{-0.1em}
    \end{minipage}

    \begin{minipage}{\columnwidth}
    \resizebox{\columnwidth}{!}{%
    \begin{tabular}
    {ccccc}
    \toprule

    {WARDEN} & 93.51$\pm$0.13 & 93.50$\pm$0.13 & 14.46$\pm$0.68 & 100.00$\pm$0.00 \\
    \midrule
    {+\gpt Attack} & 92.28$\pm$0.12 & 92.26$\pm$0.13 & 7.23$\pm$0.34 & 100.00$\pm$0.00 \\
    {+\dipper Attack} & 92.50$\pm$0.11 & 92.48$\pm$0.11 & 11.04$\pm$0.40 & 100.00$\pm$0.00 \\
    {+\nllb Attack} & 92.70$\pm$0.10 & 92.69$\pm$0.10 & 10.56$\pm$0.44 & 100.00$\pm$0.00 \\

    \bottomrule
    \end{tabular}}
    \vspace{-0.9em}
    \caption*{(d) {\agnews}}
    \vspace{-0.15em}
    \caption{The performance of \ourattack against \secondWM on \sst, \mind, \agnews, and \enron. 
    From an attacker's perspective, $\uparrow$ means higher metrics are better and $\downarrow$ means lower metrics are better. 
    } 
    \label{table:attack-performance}
    \end{minipage}
\end{table}

\reftab{table:attack-performance} presents the utility and verifiability of \secondWM \footnote{The number of watermarks, $R$, is $4$ for this experiment. Results of the impact of different $R$ are in \refapp{app:diff-R-WARDEN}.} under paraphrasing attack.\footnote{We omit the results for \firstWM here as they show similar observations; but that results are included in Appendix \reftab{table:embmarker-attack-performance}.}
In terms of utility, the paraphrasing attack only has a small negative impact on downstream performance. In terms of verifiability, for $\Delta_\text{cos}$ we see the numbers drop significantly after paraphrasing, showing that it is now harder to detect the watermark. AUC tells a similar story, with one exception: watermarks for AG News are still verifiable, suggesting the paraphrasing attack is less effective for this dataset.
We suspect this is because \agnews has much longer texts (see \reftab{table:dataset-statistics}), which means paraphrasing has the possibility of introducing new trigger words not in the original text. This is supported by our theoretical analyses (\refsec{app:attack-theoretical-proof}), which showed that although with paraphrasing we reduce the probability of higher watermark weights, at the same time this effect diminishes with longer text.
As an attacker has the freedom to select their training strategy, this means they can technically still exploit this paraphrasing vulnerability by using shorter texts when cloning the victim model.

\paragraph{Ablation Study.}
In the \refapp{sec:attack-ablation}, we present additional studies to examine the impact of various factors, such as the number of watermarks (\refapp{app:diff-R-WARDEN}), numbers of paraphrases (\refapp{sec:-attack-diff-p}), non-watermark case (\refapp{app:attack-non-wm}), attack model size (\refapp{sec:attack-diff-model-size}), and training data size (\refapp{sec:scale-up-dataset}).

\subsection{Defence Experiments}
\label{sec:defence-exps}
\begin{table}[!t]
\centering
    \begin{minipage}{0.99\columnwidth}
    \resizebox{\columnwidth}{!}{%
    \begin{tabular}
    {ccccc}
    \toprule
    {\textbf{Method}} & \textbf{ACC $\uparrow$}  & \textbf{F1 $\uparrow$}  & {\textbf{$\Delta_{cos} \uparrow$}} & {\textbf{AUC} $\uparrow$} \\
    \toprule
    {\ourdefence} & 94.58$\pm$0.21 & 94.58$\pm$0.21 & 85.67$\pm$6.92 & 100.00$\pm$0.00 \\
    \midrule
    {+\gpt Attack} & 92.73$\pm$0.25 & 92.73$\pm$0.25 & 83.58$\pm$6.43 & 100.00$\pm$0.00 \\
    {+\dipper Attack} & 91.37$\pm$0.10 & 91.36$\pm$0.10 & 83.11$\pm$6.48 & 100.00$\pm$0.00 \\
    {+\nllb Attack} & 93.24$\pm$0.24 & 93.24$\pm$0.24 & 84.28$\pm$6.04 & 100.00$\pm$0.00 \\

    \bottomrule
    \end{tabular}}
    \vspace{-0.9em}
    \caption*{(a) {\enron}}
    \vspace{-0.1em}
    \end{minipage}

    \begin{minipage}{\columnwidth}
    \resizebox{\columnwidth}{!}{%
    \begin{tabular}
    {ccccc}
    \toprule
    {\ourdefence} & 93.07$\pm$0.40 & 93.07$\pm$0.40 & 88.97$\pm$6.62 & 100.00$\pm$0.00 \\
    \midrule
    {+\gpt Attack} & 92.38$\pm$0.34 & 92.38$\pm$0.34 & 87.02$\pm$6.32 & 100.00$\pm$0.00 \\
    {+\dipper Attack} & 91.77$\pm$0.66 & 91.74$\pm$0.67 & 86.59$\pm$6.33 & 100.00$\pm$0.00 \\
    {+\nllb Attack}& 92.75$\pm$0.34 & 92.74$\pm$0.34 & 87.78$\pm$6.27 & 100.00$\pm$0.00 \\

    \bottomrule
    \end{tabular}}
    \vspace{-0.9em}
    \caption*{(b) {\sst}}
    \vspace{-0.1em}
    \end{minipage}

    \begin{minipage}{\columnwidth}
    \resizebox{\columnwidth}{!}{%
    \begin{tabular}
    {ccccc}
    \toprule
    {\ourdefence} & 77.11$\pm$0.08 & 51.03$\pm$0.26 & 87.74$\pm$6.17 & 100.00$\pm$0.00 \\
    \midrule
    {+\gpt Attack} & 76.72$\pm$0.05 & 50.62$\pm$0.25 & 87.44$\pm$6.17 & 100.00$\pm$0.00 \\
    {+\dipper Attack} & 76.58$\pm$0.08 & 49.99$\pm$0.23 & 86.81$\pm$5.90 & 100.00$\pm$0.00 \\
    {+\nllb Attack} & 76.47$\pm$0.14 & 49.85$\pm$0.26 & 87.54$\pm$5.91 & 100.00$\pm$0.00 \\

    \bottomrule
    \end{tabular}}
    \vspace{-0.9em}
    \caption*{(c) {\mind}}
    \vspace{-0.1em}
    \end{minipage}

    \begin{minipage}{\columnwidth}
    \resizebox{\columnwidth}{!}{%
    \begin{tabular}
    {ccccc}
    \toprule
    {\ourdefence} & 93.15$\pm$0.08 & 93.14$\pm$0.08 & 88.35$\pm$6.6 & 100.00$\pm$0.00 \\
    \midrule
    {+\gpt Attack} & 92.22$\pm$0.10 & 92.20$\pm$0.10 & 88.02$\pm$6.14 & 100.00$\pm$0.00 \\
    {+\dipper Attack} & 92.46$\pm$0.18 & 92.45$\pm$0.18 & 87.79$\pm$6.14 & 100.00$\pm$0.00 \\
    {+\nllb Attack} & 92.43$\pm$0.08 & 92.42$\pm$0.08 & 88.44$\pm$5.91 & 100.00$\pm$0.00 \\

    \bottomrule
    \end{tabular}}
    \vspace{-0.9em}
    \caption*{(d) {\agnews}}
    \vspace{-0.15em}
    \end{minipage}
    \caption{The performance of \ourdefence watermark for different scenarios on SST2, MIND, AG News, and Enron datasets. 
    From a defender's perspective, $\uparrow$ means higher metrics are better. 
    All the metrics are in \%.} 
    \label{table:defence-perf}
\end{table}

\paragraph{Watermark Performance.}
We present the utility and verifiability performance of \ourdefence against paraphrasing attacks in \reftab{table:defence-perf}. If we compare \ourdefence to \secondWM in \reftab{table:attack-performance}, their downstream performance is about the same, suggesting they are all competitive in terms of maintaining utility. However, \ourdefence is better when it comes to verifiability, as its AUC is 100\% in all cases. Examining the impact of the paraphrasing attack, \ourdefence is a clear winner here, as all verifiability metrics see minimal changes (most importantly, AUC is still 100\%). These results empirically validate that \ourdefence is not susceptible to paraphrasing attacks.
We now present additional analyses to understand the impact of hyper-parameters $k$ and $w$. For these, we only look at utility performance as verifiability does not change based on these hyper-parameters.

\begin{figure}[!t]
    \centering
    \includegraphics[width=0.95\linewidth]{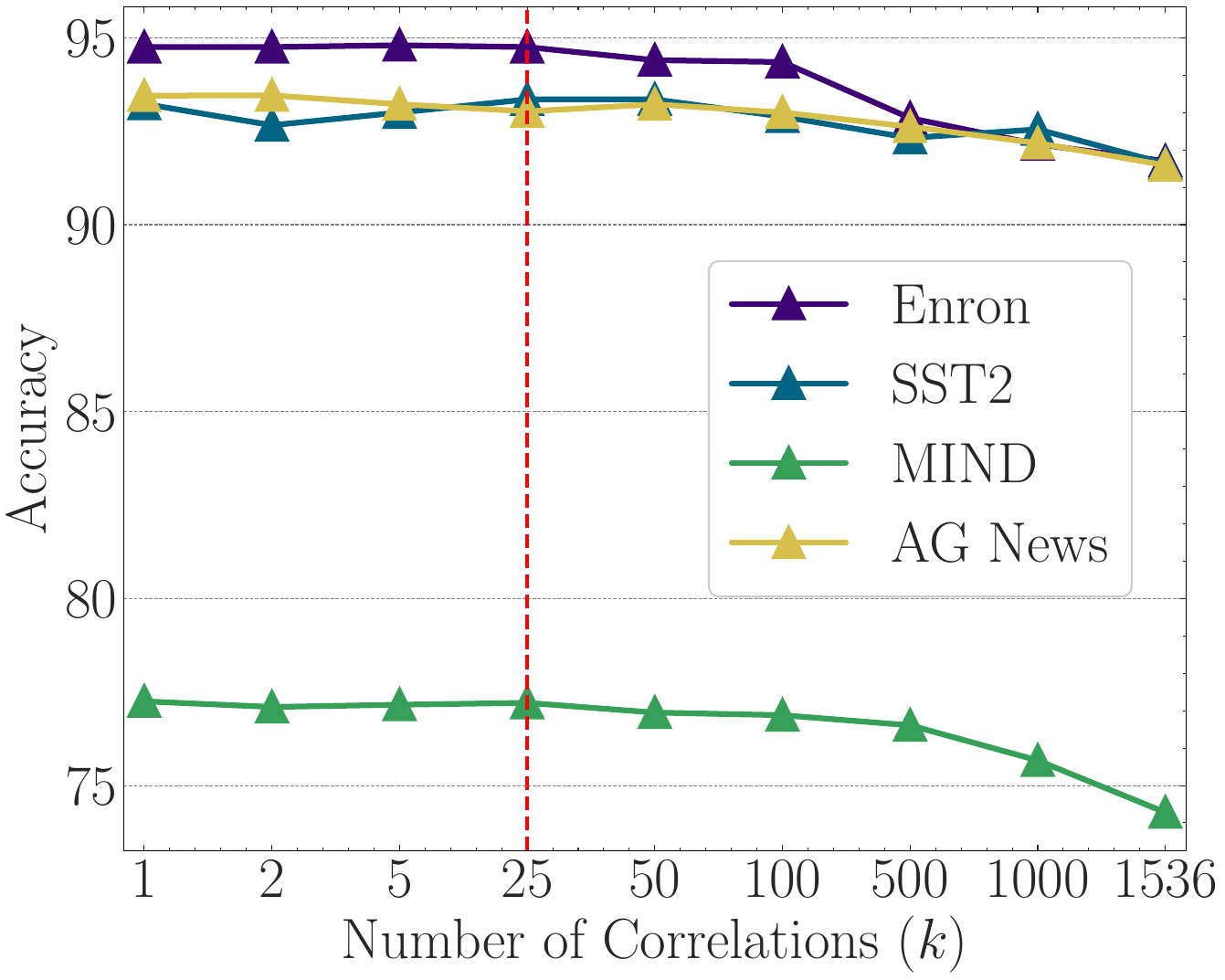}
    \caption{Impact of different values of correlations ($k$) on watermark utility. We ignore verifiability as they are always perfect (\ie 100\%). The \textcolor{red}{red} vertical dashed line represents our chosen value ($k=25$).}
    \label{fig:diff-k-WET}
\end{figure}
\paragraph{Number of correlations ($k$).}
\label{sec:diff-k-defence}

In \reffig{fig:diff-k-WET}, we can see that for higher values of $k$ ($>100$), we start seeing degradation in the watermarked embedding utility. When we consider more original embedding dimensions for calculating watermarked embedding, the increased complexity introduces confusion, making it harder for the surrogate model to learn the underlying semantic properties of the embeddings. 
Hence, we chose $k=25$ in our experiments. A more comprehensive table with full results is provided in Appendix \reftab{table:diff-k-defence}.

\begin{figure}[h]
    \centering
    \includegraphics[width=0.95\linewidth]{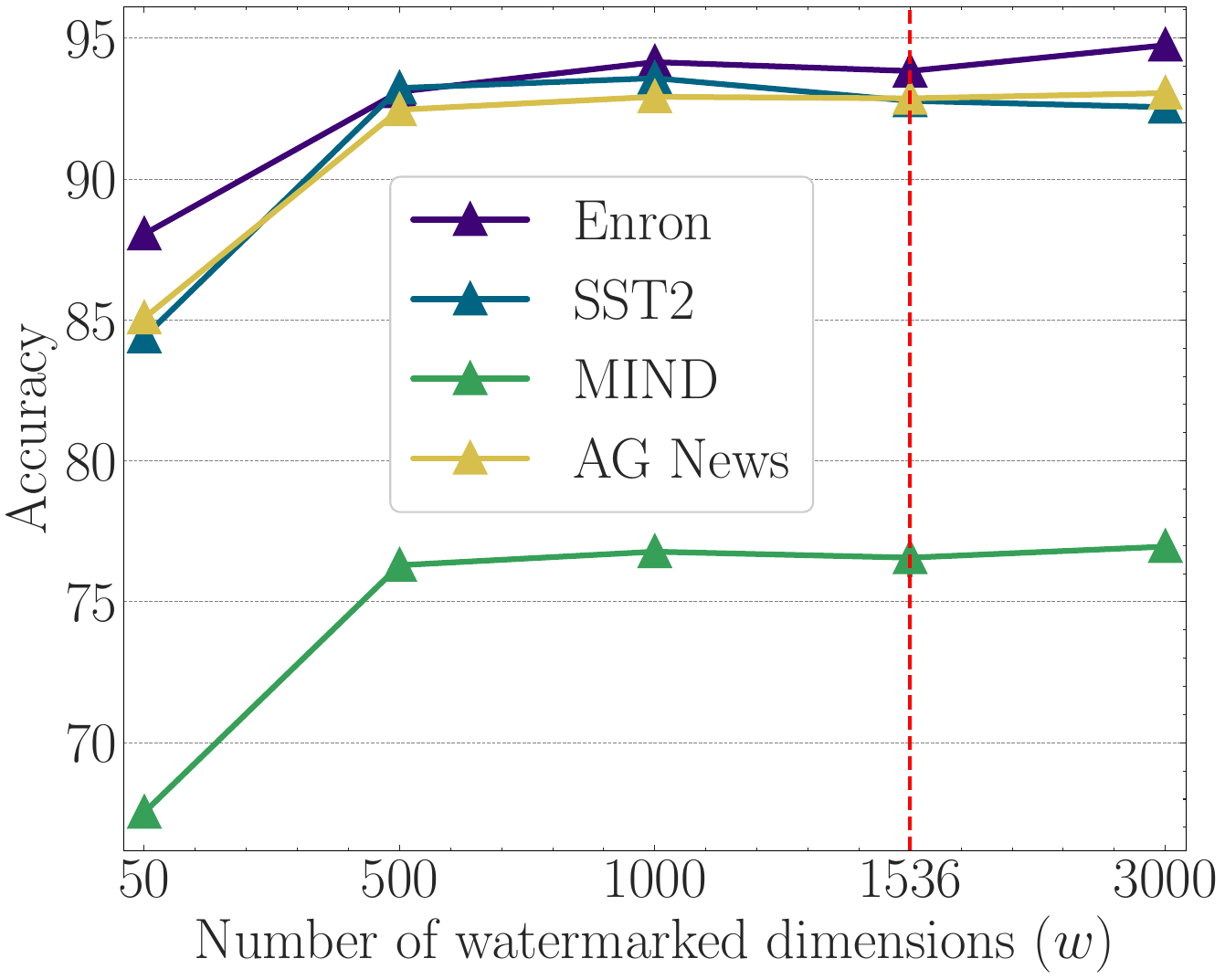}
    \caption{Impact of different values of watermark dimensions ($w$) on watermark utility. The \textcolor{red}{red} vertical dashed line represents our chosen value  ($w=1536$).}
    \label{fig:diff-w-WET}
\end{figure}
\paragraph{Number of watermarked dimensions ($w$).}
\label{sec:diff-h-defence}
We can observe from \reffig{fig:diff-w-WET} that smaller values might also work. 
This demonstrates another benefit of our \ourdefence technique: it can also be used for compression. That said, utility is only measured using simple classification tasks (following prior studies) and as such these results may be different if the embeddings are used for more complex tasks. As such, we use $w=n$ in our experiments.
For more results on different values of $w$, see  Appendix \reftab{table:diff-h-defence}.

\paragraph{Ablation Study.} In our watermark verification process, which includes reverse transformation, we evaluate the resilience of \ourdefence to perturbations. Our findings show that \ourdefence remains verifiable even under significant utility loss, highlighting its robustness; see \refapp{app:gauss-noise-defence} for results. In \refapp{app:diff-verif-size-defence}, we demonstrate that \ourdefence requires very few samples (even one) for watermark verification; a contrast to EmbMarker and WARDEN which require multiple samples for verification. In \refapp{app:diff-matrices} we present additional results with different configurations of the transformation matrix, a critical component of \ourdefence. Lastly, in \refapp{app:diff-model-size-defence} we show that \ourdefence is not affected by the attack model size.

\section{Conclusion}
We highlight the vulnerabilities of existing EaaS watermarks against paraphrasing in an imitation attack. Our approach involves generating multiple paraphrases and combining their embeddings, which effectively reduces the impact of trigger words and thereby removes the watermark. To address this shortcoming, we devise a simple watermarking technique, \ourdefence, which applies linear transformations to the original embeddings to generate watermarked embeddings. Our experiments demonstrate that \ourdefence is robust against paraphrasing attacks and has a much stronger verifiability performance. Additionally, we conduct ablation studies to assess the contribution of each component in the paraphrasing attack and \ourdefence.

\section*{Limitations}

With the current design of the circulant transformation matrix, the matrix is compromised if an attacker manages to recover any single row in the matrix. A better approach could be to use different weights (more in \refapp{app:diff-matrices}) for each row in the circulant matrix, but this means we would lose crucial properties such as invertibility and full rank. Therefore, we opted to retain the current design, though we acknowledge that the design can be potentially further improved.

For utility, we focus on simple classification tasks in line with existing studies; however, these tasks may not be sufficient to fully validate embedding quality. Moving forward, we believe it is important we start exploring other more complex NLP tasks, such as retrieval and generation, to gain a deeper understanding of the true impact of introducing watermarks into embeddings.

\section*{Ethics Statement}

We introduce paraphrasing as a new form of attack against EaaS watermarks. We want to clarify that our intention here is to raise awareness about this new form of attack, as we believe the first step in improving security is by exposing vulnerabilities. As a countermeasure, we therefore also introduce a new watermarking technique, \ourdefence, that is resilient against paraphrasing attacks.

\section*{Acknowledgement}
We would like to appreciate the valuable feedback from all anonymous reviewers. This research was supported by The University of Melbourne’s Research Computing Services and the Petascale Campus Initiative.
This project is supported by the Australian Research Council Linkage Project (ID: LP210200917).
Q.~Xu acknowledges support from 2024 FSE Strategic Startup and FSE Travel Grant. A.~Shetty acknowledges support from Avashya Foundation.

\bibliography{custom}

\clearpage
\appendix
\section*{Appendix}

\section{Experimental Settings}
\label{sec:app-exp-setting}

\subsection{Prompt-Based Paraphrasing Attacks}
\label{app:gpt-exp-setting}
We utilise the prompts chosen through the below analysis and query the \texttt{gpt-3.5-turbo-0125} model using the maximum token length of 1000 and temperature of 0.7. One must note that for all paraphrasing attack setups, it is up to $P$ paraphrases. We query the models to generate $P$ paraphrases; however, we do not make repeated queries to ensure we have $P$ paraphrases for cases where it generates fewer paraphrases. 
\paragraph{Prompts.}
We evaluated two prompts:
\paragraph{\texttt{PROMPT 1} \citep{kirchenbauer2024on}:} \textit{``As an expert copy-editor, please rewrite the following text in your own voice while ensuring that the final output contains the same information as the original text and has roughly the same length. Please paraphrase all sentences and do not omit any crucial details. Additionally, please take care to provide any relevant information about public figures, organisations, or other entities mentioned in the text to avoid any potential misunderstandings or biases.''}
\paragraph{\texttt{PROMPT 2} \cite{RTT-wm}:} \textit{``You are a helpful assistant to rewrite the text. Rewrite the following text:''}

We use \texttt{PROMPT 2} in our experiments unless stated. Performance of \texttt{PROMPT 1} was evaluated for \enron, it was subpar compared to \texttt{PROMPT 2}. It was because \texttt{PROMPT 1} could explain and expand on short (few or single words) input text. This leads to a significant deviation from the original text. Moreover, increases the chances of incorporating the trigger words.

\paragraph{Attack Cost.} The approximate total number of tokens for all the datasets is \enron (377K), \sst (1M), \mind (2M), and \agnews (7M). Considering $P=5$ and assuming similar tokens in the output, the expected cost of generating paraphrases using \gpt (input - \$0.50 / 1M tokens and output - \$1.50 / 1M  tokens) would be just under \$105.

\subsection{DIPPER Paraphrasing Attacks}
\label{app:dipper-exp-setting}
We employ DIPPER \citep{krishna2024paraphrasing}, an explicitly trained paraphraser with hyperparameters (\texttt{lex} and \texttt{div}), to control the paraphrasing quality. 
As per findings in \citet{krishna2024paraphrasing}, DIPPER performs at par with GPT-3.5 models in terms of controlling the diversity and quality of paraphrases.
We adopt a moderate setting for all our experiments: \texttt{lex} = 40 and \texttt{div} = 40. It still ensures significant changes to the text but, at the same time, maintains a high quality of paraphrases.

\subsection{Round-Trip Translation Paraphrasing Attacks}
\label{app:nllb-exp-setting}
\begin{table}[ht]
    \begin{minipage}{0.99\columnwidth}
    \resizebox{\textwidth}{!}{%
    \centering
    \begin{tabular}{cccc}

        \toprule
        {Language} & {IDO 639-I} & {ISO 639-2/T} & {Language family} \\
        \midrule
        {Chinese (Simpl)} & zh & zho\_simpl & Sino-Tibetan \\
        {Japanese} &  ja & jpn & Other \\
        {French} &  fr & fra & Indo-European-Romance \\
        {German} & de & deu & Indo-European-Germanic \\
        {Hindi} &  hi & hin & Indo-European-Indo-Aryan \\
        \bottomrule
    \end{tabular}}
    \caption{For each language we use in RTT, we list its language name, ISO code and language family \citep{zhu-etal-2024-multilingual}.}
    \label{table:RTT-lang-info}
    \end{minipage}
\end{table}

\textbf{Round-trip translation} (RTT) involves translating text to another language and then back-translating to the original language (\eg English $\rightarrow$ German $\rightarrow$ English). It is commonly used for evaluating machine translation systems because the original and resulting text could vary significantly \citep{somers-2005-round}.  We explore translations (represented in \reftab{table:RTT-lang-info}) for languages considerably different from English (our original language), such as Chinese, German, and others, covering a diverse group such that the translated text will probably have more modifications.  \gpt is still not SOTA for multilingual translations, as found in \citet{zhu-etal-2024-multilingual}. Hence, we use the 1.3B NLLB model variant, an open-source multilingual model.

\subsection{Baseline Method Details and Hyperparameters}
For fair comparisons, we use the original default settings of the baseline methods unless specified otherwise.

\paragraph{EmbMarker.} The size of the trigger word set is 20, and the maximum number of trigger words m is 4, with a frequency interval for trigger words of [0.5\%, 1\%]. We use BERT \citep{devlin-etal-2019-bert} as the backbone, with a two-layer feed-forward network for imitation attacks and a mean squared error (MSE) loss for training.

\paragraph{WARDEN.} The settings remain the same as described for EmbMarker above, with the number of watermarks ($R$) set to 4.

\subsection{Definition of Positive and Negative Test Samples for Watermark Verification}
\label{sec:pos-neg-samples}
These are used for the calculation of AUC metrics.

\paragraph{``Positive'' Samples.} In all the \ourdefence experiments, these are (copied) embeddings $E_w$ returned by the copied model $\mathbb{S}_{v}$ that should be classified as watermarked embeddings. Whereas, for \ourattack experiments, these are embeddings from the backdoor (all trigger words) verification dataset.

\paragraph{``Negative'' Samples.} In all the \ourdefence experiments, these are contrast (copied) embeddings $E_c$ returned by another copied model $\mathbb{S}_{v}^*$. We train another copied model following a similar process using a different transformation matrix and use this model's embeddings as the non-watermarked embeddings to make it more challenging. Similarly, in \ourattack experiments, these are embeddings from the benign (no trigger words) verification dataset.

\subsection{Code and Compute Details}
We expand on the watermarking implementation by \citet{shetty2024warden}. We make extensive use of the Huggingface Transformers \citep{wolf-etal-2020-transformers} framework and AdamW \citep{loshchilovdecoupled} for models and datasets library \citep{lhoest-etal-2021-datasets} for data assessed in this work.
To spur future research in this area, we intend to make the embeddings and code available post-acceptance.

All experiments were conducted using a single A100 GPU with CUDA 11.7 and PyTorch 2.1.2. To ensure that the impact of the watermarking technique is isolated from other variables, we assume that both the victim model and imitators utilize the same datasets. Additionally, we presume that the extracted model is trained solely on the watermarked outputs of the victim model.

\section{Paraphrasing Attack Analyses}
\label{sec:attack-ablation}
\begin{table}[!ht]
\centering
    \begin{minipage}{0.99\columnwidth}
    \resizebox{\columnwidth}{!}{%
    \begin{tabular}
    {ccccc}
    \toprule
    {\textbf{Method}} & \textbf{ACC $\uparrow$}  & \textbf{F1 $\uparrow$}  & {\textbf{$\Delta_{cos} \downarrow$}} & {\textbf{AUC} $\downarrow$} \\
    \toprule
    {EmbMarker} & 94.58$\pm$0.09 & 94.58$\pm$0.09 & 5.44$\pm$0.13 & 93.50$\pm$0.97 \\
    \midrule
    {+\gpt Attack} & 92.80$\pm$0.19 & 92.80$\pm$0.19 & -0.03$\pm$0.07 & 49.80$\pm$1.35 \\
    {+\dipper Attack} & 92.35$\pm$0.48 & 92.35$\pm$0.49 & 0.63$\pm$0.16 & 61.85$\pm$4.52 \\
    {+\nllb Attack} & 93.38$\pm$0.20 & 93.38$\pm$0.20 & 0.69$\pm$0.20 & 65.25$\pm$3.68 \\
    
    \bottomrule
    
    \end{tabular}}
    \vspace{-0.9em}
    \caption*{(a) {\enron}}
    \vspace{-0.1em}
    \end{minipage}

    \begin{minipage}{\columnwidth}
    \resizebox{\columnwidth}{!}{%
    \begin{tabular}
    {ccccc}
    \toprule
    {EmbMarker} & 92.89$\pm$0.25 & 92.89$\pm$0.25 & 4.05$\pm$2.70 & 95.04$\pm$2.30 \\
    \midrule
    {+\gpt Attack} & 92.86$\pm$0.17 & 92.86$\pm$0.17 & 0.68$\pm$0.10 & 68.20$\pm$2.94 \\
    {+\dipper Attack} & 91.31$\pm$0.24 & 91.27$\pm$0.25 & 0.94$\pm$0.12 & 79.95$\pm$3.89 \\
    {+\nllb Attack} & 92.66$\pm$0.55 & 92.64$\pm$0.55 & 0.76$\pm$0.11 & 78.20$\pm$3.60 \\

    \bottomrule
    \end{tabular}}
    \vspace{-0.9em}
    \caption*{(b) {\sst}}
    \vspace{-0.1em}
    \end{minipage}
    
    \begin{minipage}{\columnwidth}
    \resizebox{\columnwidth}{!}{%
    \begin{tabular}
    {ccccc}
    \toprule
    {EmbMarker} & 77.34$\pm$0.06 & 51.63$\pm$0.16 & 3.93$\pm$0.11 & 93.10$\pm$0.94 \\
    \midrule
    {+\gpt Attack} & 77.01$\pm$0.07 & 51.23$\pm$0.13 & 1.04$\pm$0.08 & 67.75$\pm$1.66 \\
    {+\dipper Attack} & 76.83$\pm$0.09 & 50.56$\pm$0.11 & 2.22$\pm$0.09 & 90.15$\pm$1.68 \\
    {+\nllb Attack} & 76.59$\pm$0.14 & 50.32$\pm$0.26 & 2.11$\pm$0.07 & 85.80$\pm$1.34 \\

    \bottomrule
    \end{tabular}}
    \vspace{-0.9em}
    \caption*{(c) {\mind}}
    \vspace{-0.1em}
    \end{minipage}

    \begin{minipage}{\columnwidth}
    \resizebox{\columnwidth}{!}{%
    \begin{tabular}
    {ccccc}
    \toprule
    {EmbMarker} & 93.47$\pm$0.12 & 93.47$\pm$0.12 & 12.53$\pm$0.67 & 100.00$\pm$0.00 \\
    \midrule
    {+\gpt Attack} & 92.17$\pm$0.04 & 92.15$\pm$0.04 & 4.66$\pm$0.36 & 99.15$\pm$0.34 \\
    {+\dipper Attack} & 92.47$\pm$0.10 & 92.45$\pm$0.10 & 6.68$\pm$0.40 & 100.00$\pm$0.00 \\
    {+\nllb Attack} & 92.76$\pm$0.13 & 92.74$\pm$0.13 & 6.3$\pm$0.35 & 100.00$\pm$0.00 \\

    \bottomrule
    \end{tabular}}
    \vspace{-0.9em}
    \caption*{(d) {\agnews}}
    \vspace{-0.1em}
    \caption{The performance of \ourattack against \firstWM for different scenarios, similar to \reftab{table:attack-performance}. 
    } 
    \label{table:embmarker-attack-performance}
    \end{minipage}
\end{table}

In this section, we perform detailed analysis and ablation studies for \ourattack.

\subsection{Analysis of Watermarking Weight Distribution after Paraphrasing}
\label{app:attack-theoretical-proof}

\begin{figure*}[p]
    \centering
    \begin{subfigure}{0.47\textwidth}
    \centering\includegraphics[width=\linewidth,keepaspectratio]{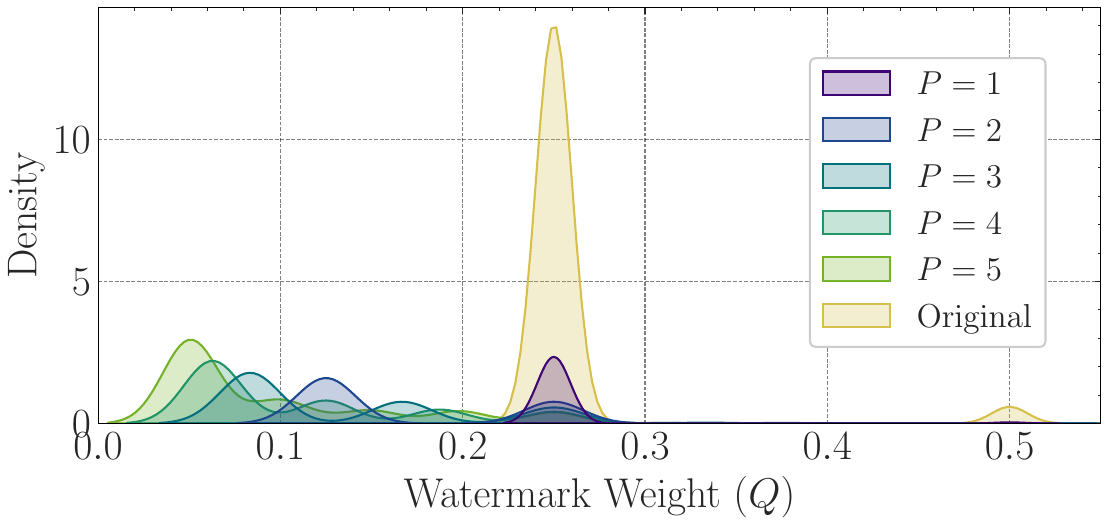}
    \caption{\enron}
    \end{subfigure}
    \begin{subfigure}{0.48\textwidth}
    \centering\includegraphics[width=\linewidth,keepaspectratio]{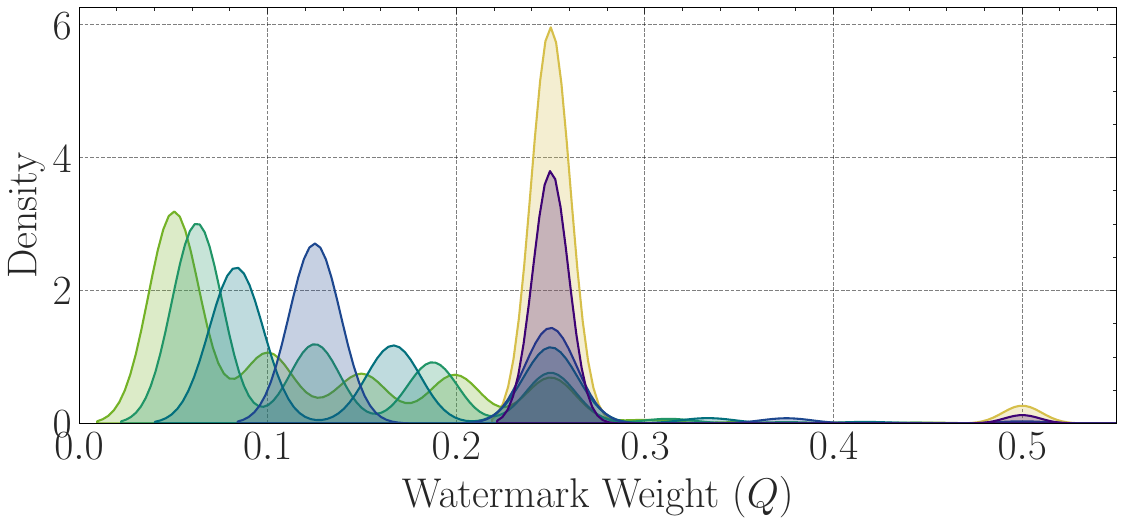}
    \caption{\sst}
    \end{subfigure}
    \begin{subfigure}{0.48\textwidth}
    \centering\includegraphics[width=\linewidth,keepaspectratio]{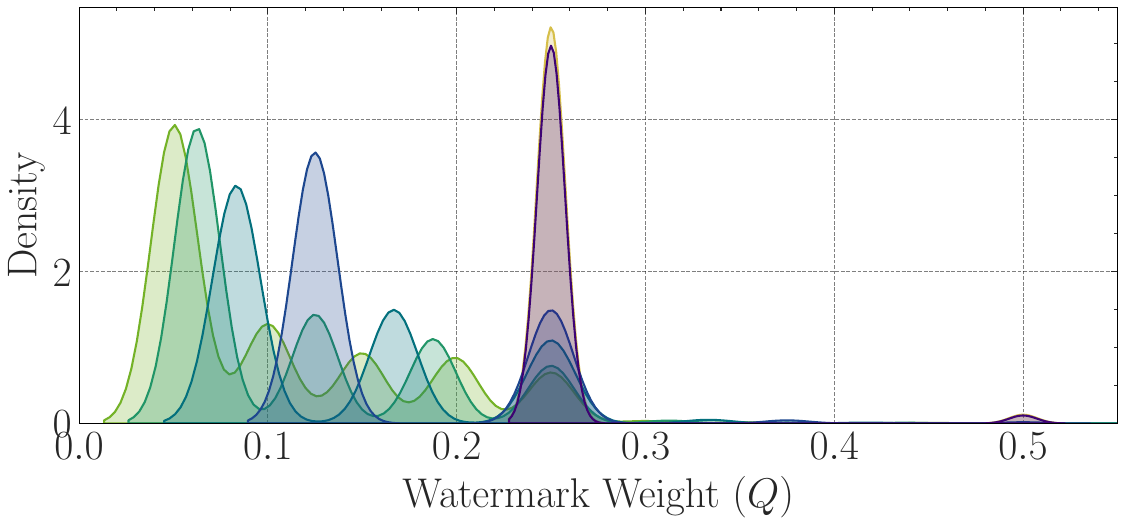}
    \caption{\mind}
    \end{subfigure}
    \begin{subfigure}{0.48\textwidth}
    \centering\includegraphics[width=\linewidth,keepaspectratio]{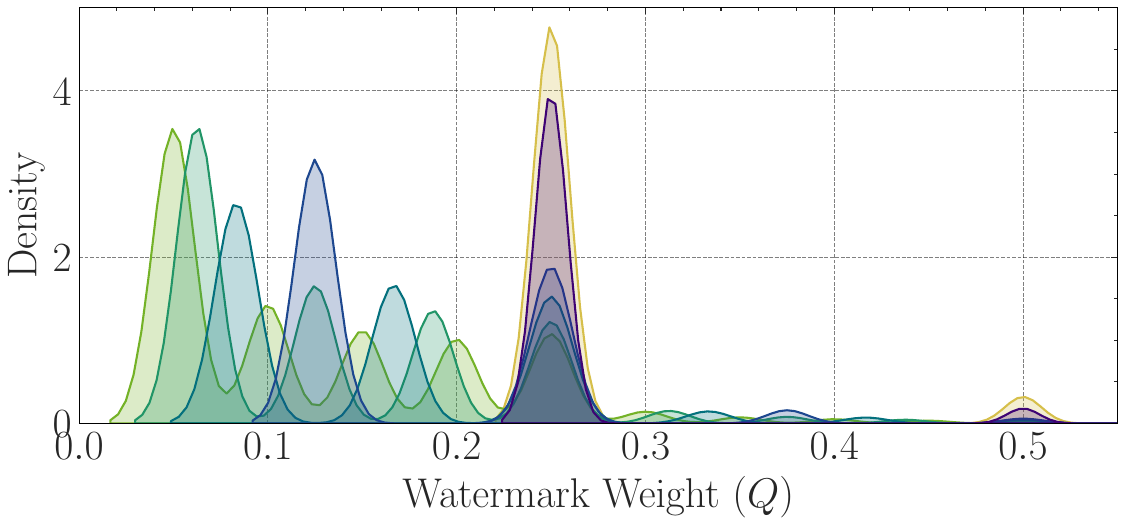}
    \caption{\agnews}
    \end{subfigure}
    \caption{The analysis of watermark weight on different datasets using NLLB paraphrases. We queried NLLB up to $P=5$ times to produce paraphrases.}

    \label{fig:nllb-wm-wt-analysis}
\end{figure*}

\begin{figure*}[p]
    \centering
    \begin{subfigure}{0.473\textwidth}
    \centering\includegraphics[width=\linewidth,keepaspectratio]{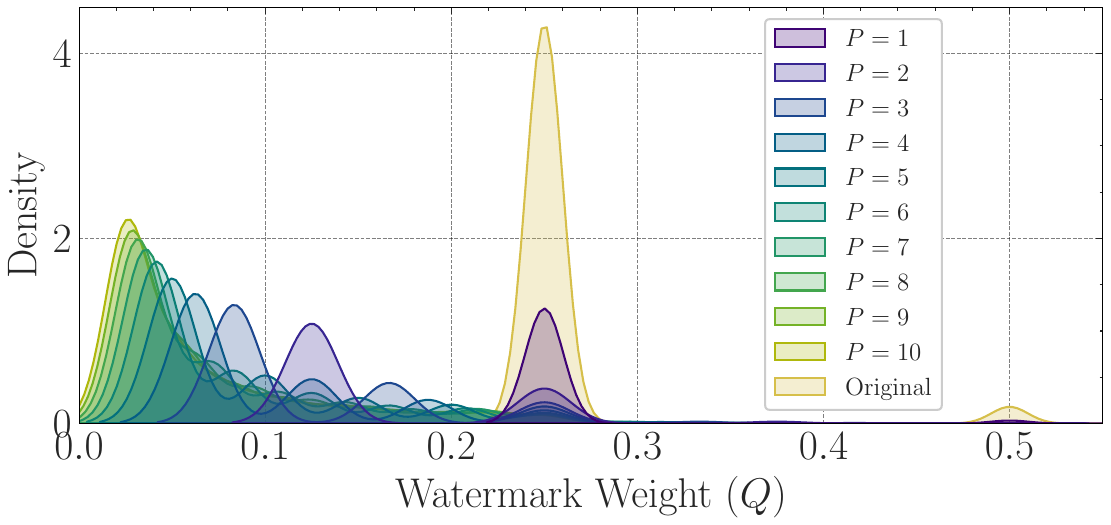}
    \caption{\enron}
    \end{subfigure}
    \begin{subfigure}{0.48\textwidth}
    \centering\includegraphics[width=\linewidth,keepaspectratio]{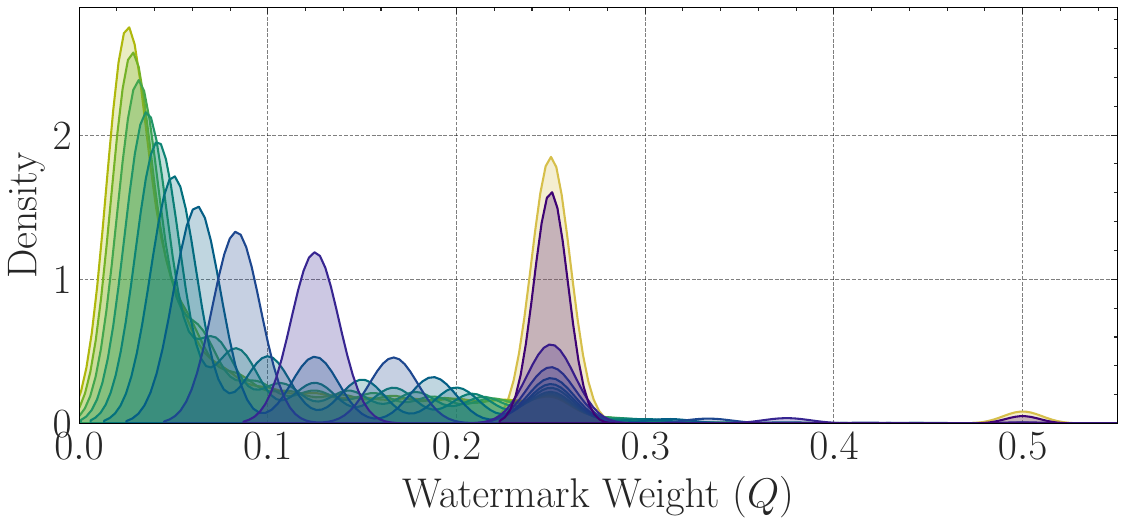}
    \caption{\sst}
    \end{subfigure}
    \begin{subfigure}{0.473\textwidth}
    \centering\includegraphics[width=\linewidth,keepaspectratio]{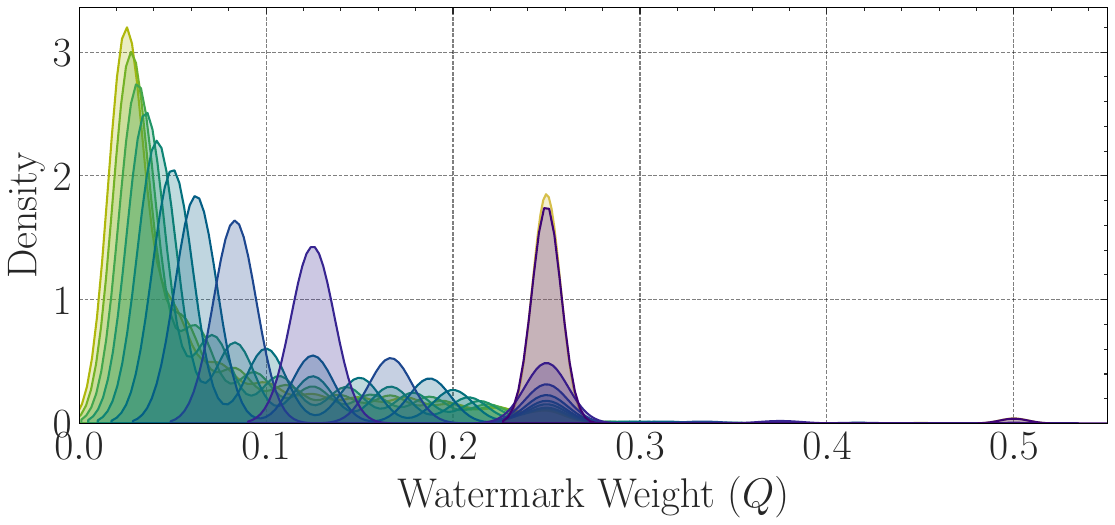}
    \caption{\mind}
    \end{subfigure}
    \begin{subfigure}{0.48\textwidth}
    \centering\includegraphics[width=\linewidth,keepaspectratio]{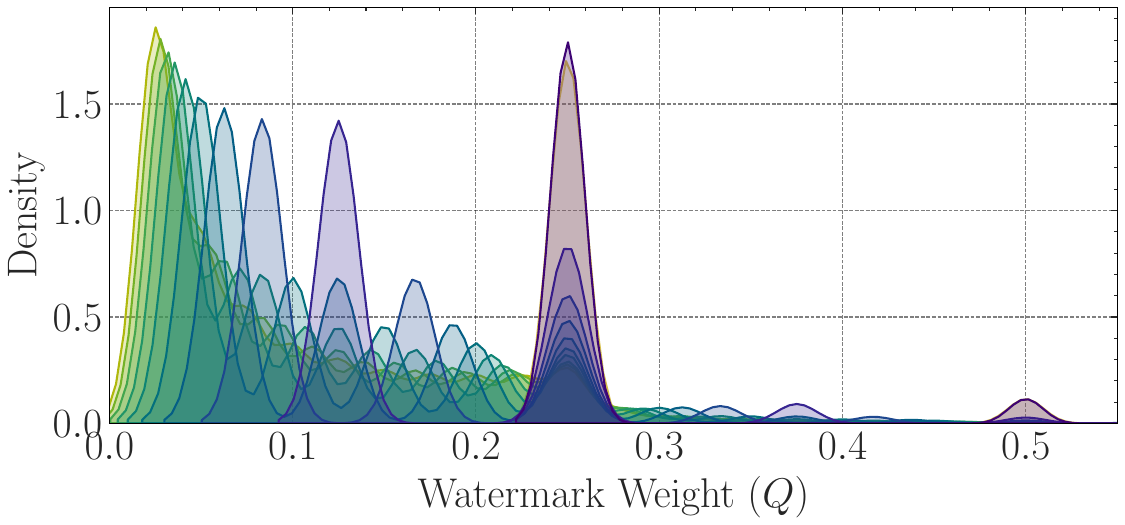}
    \caption{\agnews}
    \end{subfigure}
    \caption{The analysis of watermark weight on different datasets using DIPPER paraphrases. We queried DIPPER up to $P=10$ times to produce paraphrases.}
    \label{fig:dipper-wm-wt-analysis}
\end{figure*}

We analyse the impact of paraphrasing on watermarking weights in a simplified setting as follows:
\begin{figure}[h!]
    \centering
    \includegraphics[width=0.9\linewidth,keepaspectratio]{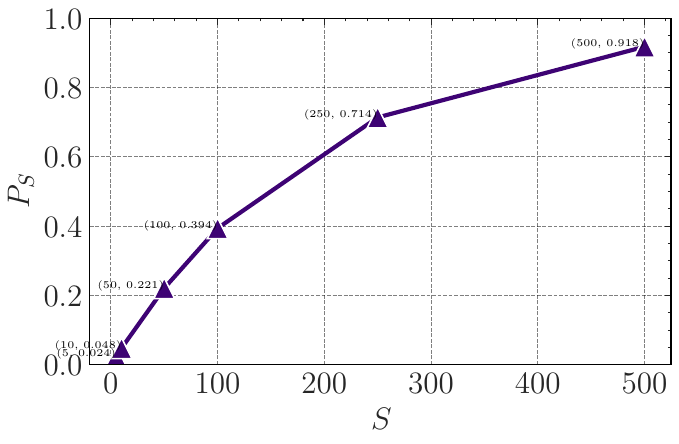}
    \caption{$\mathbb{P}_S$ for different values of sentence length $|S|$. With increasing text length (S), there are higher chances of trigger words in the text $\mathbb{P}_S$.}
    \label{fig:diff-S-Pts}
\end{figure}
\begin{itemize}
    \item Each token has low probability $\mathbb{P}_t$ of being in the trigger words set $t$.
    \item Sentences with equal or more than one trigger acquire the same watermark weight $\lambda > 0$.
    \item Average of $P$ paraphrase sentences gives watermark weight $\lambda \cdot Q_{P}$ and single sentence gives watermark weight $\lambda \cdot Q_{S}$.
\end{itemize}

As per the above assumptions, the probability of a sentence $S$ having trigger words is $$\mathbb{P}_S = 1 - ( 1 - \mathbb{P}_t)^{|S|}.$$

The weight by a single sentence is $\lambda \cdot Q_{S}$, where

    \begin{align}
        Q_{S} &\sim \text{Bernoulli}(\mathbb{P}_S).
    \end{align}

The weight by averaged paraphrasing embeddings are equivalent to $\lambda \cdot Q_{P}$, where
    \begin{align}
        Q_{P} &= \frac{X_{P}}{P},
        X_{P} = \sum_{i=1}^{P}Q_S^{i}, \\ 
        X_{P} &\sim \text{Binomial}(P, \mathbb{P}_S).
    \end{align}

As per \secondWM setting used, trigger word frequency is $[ 0.5\%, 1\% ]$. Therefore, assuming a generic case, $\mathbb{P}_t = 0.005$ and $|S| = 50$ (refer to \reftab{table:dataset-statistics}), $\mathbb{P}_S = 0.222$.

When $P=10$, $\mathbb{P}(Q_S > a) > \mathbb{P} (Q_{P} > a)$ for all $a > 0.3$. Similarly, when $P=5$, $\mathbb{P}(Q_S > a) > \mathbb{P} (Q_{P} > a)$ for all $a > 0.4$. 
This demonstrates that paraphrasing will give the attackers a higher chance of getting watermark weights lower than a low threshold. As a result, watermarks will be diluted and neglected in the training for imitation attacks.
However, this diminishes with increasing text length ($|S|$), as observed in \agnews dataset in \refsec{sec:attack-perf}.

\subsection{Number of Watermarks ($R$) in \secondWM}
\label{app:diff-R-WARDEN}
\begin{figure*}[h]
    \centering
    \begin{subfigure}{0.25\textwidth}
    \centering\includegraphics[width=\linewidth,keepaspectratio]{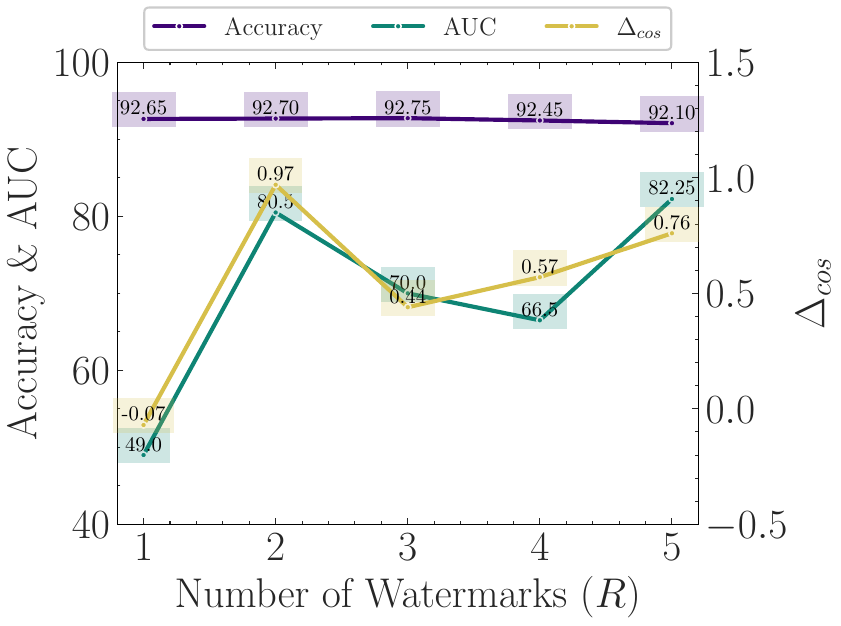}
    \caption{\enron}
    \end{subfigure}
    \begin{subfigure}{0.23\textwidth}
    \centering\includegraphics[width=\linewidth,keepaspectratio]{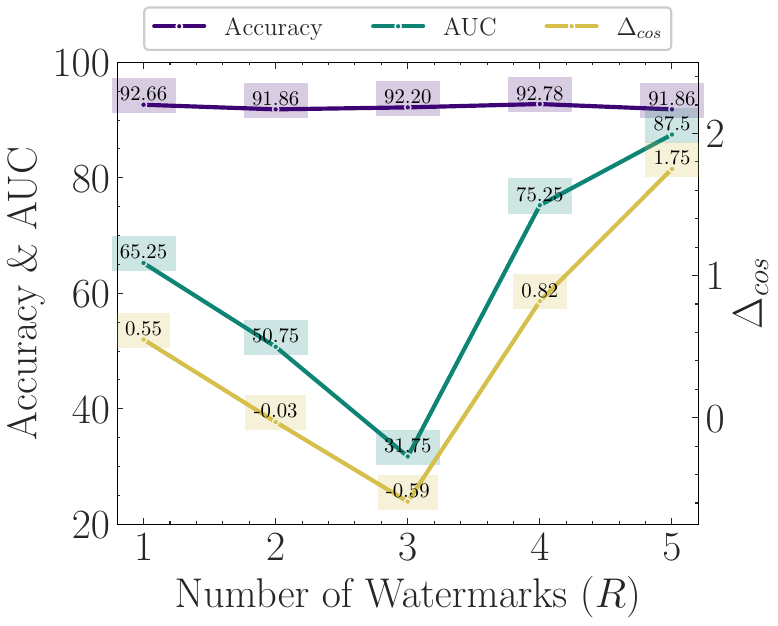}    \caption{\sst}
    \end{subfigure}
    \begin{subfigure}{0.23\textwidth}
    \centering\includegraphics[width=\linewidth,keepaspectratio]{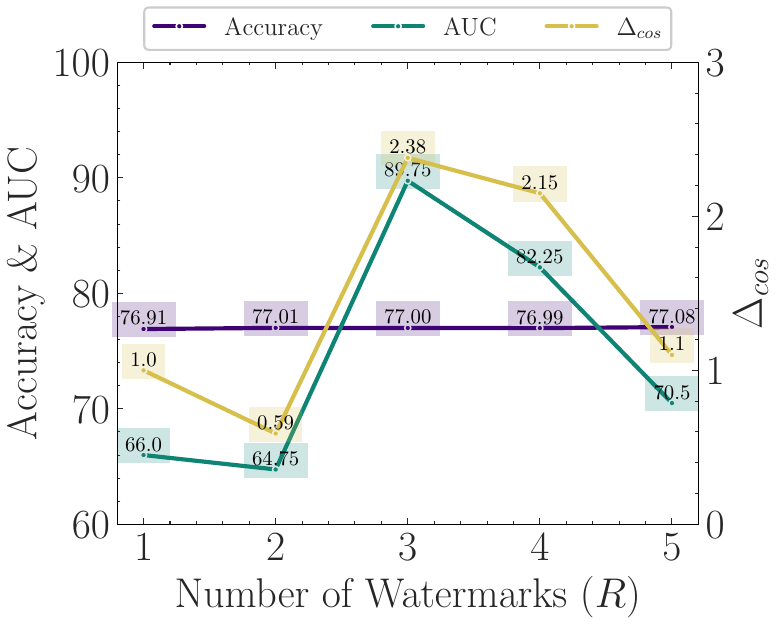}    \caption{\mind}
    \end{subfigure}
    \begin{subfigure}{0.245\textwidth}
    \centering\includegraphics[width=\linewidth,keepaspectratio]{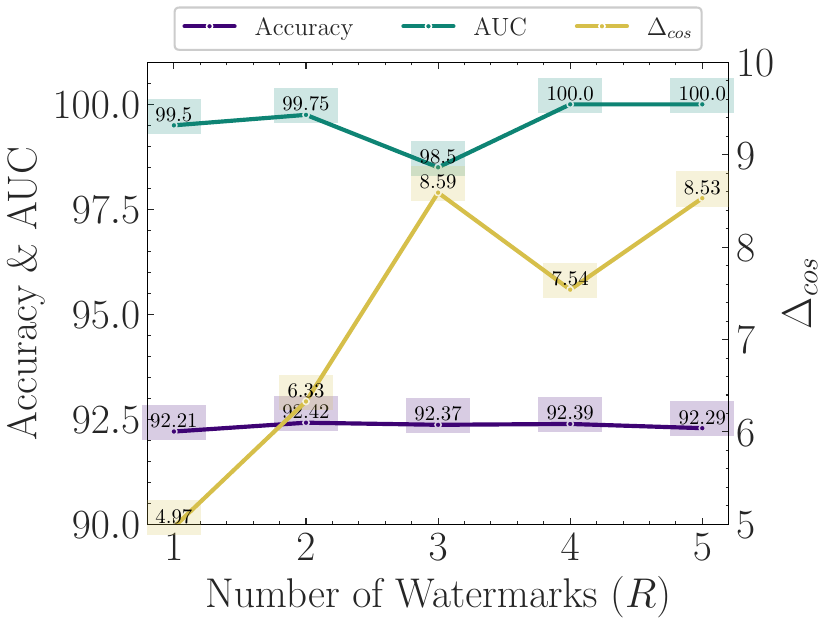}    \caption{\agnews}
    \end{subfigure}
    \caption{\gpt paraphrase attack performance different number of watermarks ($R$) for all the datasets.}
    \label{fig:diff-R-WARDEN-attack-perf}
\end{figure*}

As expected, watermark verification performance (green and yellow lines) shows an upward trend with stable watermark utility (blue line) as shown in \reffig{fig:diff-R-WARDEN-attack-perf}.

\subsection{Number of Paraphrases ($P$)}
\label{sec:-attack-diff-p}

\begin{figure*}[h!]
    \centering
    \begin{subfigure}{0.24\textwidth}
    \centering\includegraphics[width=\linewidth,keepaspectratio]{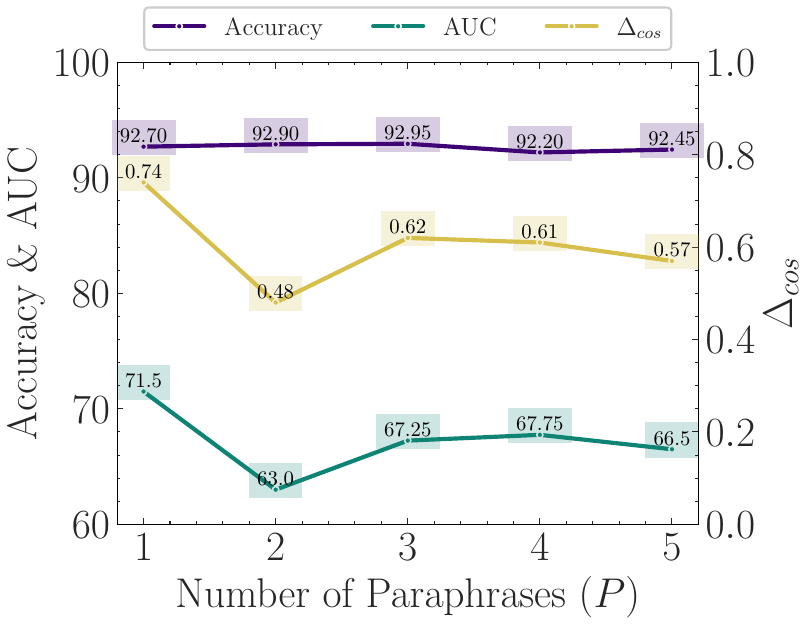}
    \caption{\enron}
    \end{subfigure}
    \begin{subfigure}{0.24\textwidth}
    \centering\includegraphics[width=\linewidth,keepaspectratio]{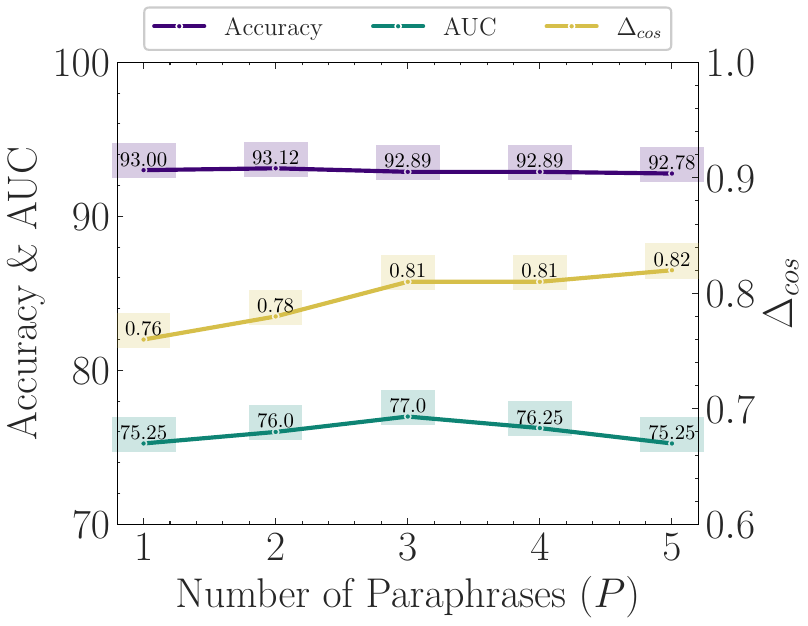}
    \caption{\sst}
    \end{subfigure}
    \begin{subfigure}{0.235\textwidth}
    \centering\includegraphics[width=\linewidth,keepaspectratio]{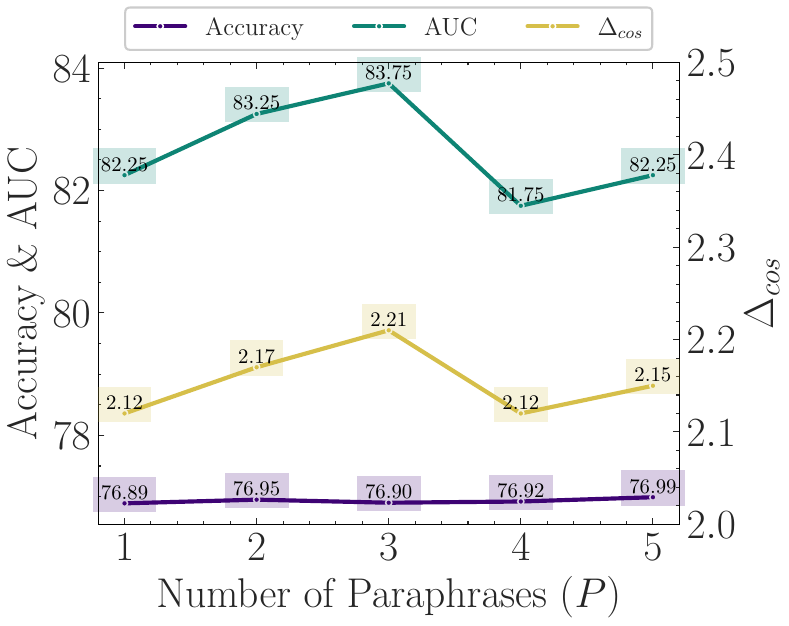}
    \caption{\mind}
    \end{subfigure}
    \begin{subfigure}{0.24\textwidth}
    \centering\includegraphics[width=\linewidth,keepaspectratio]{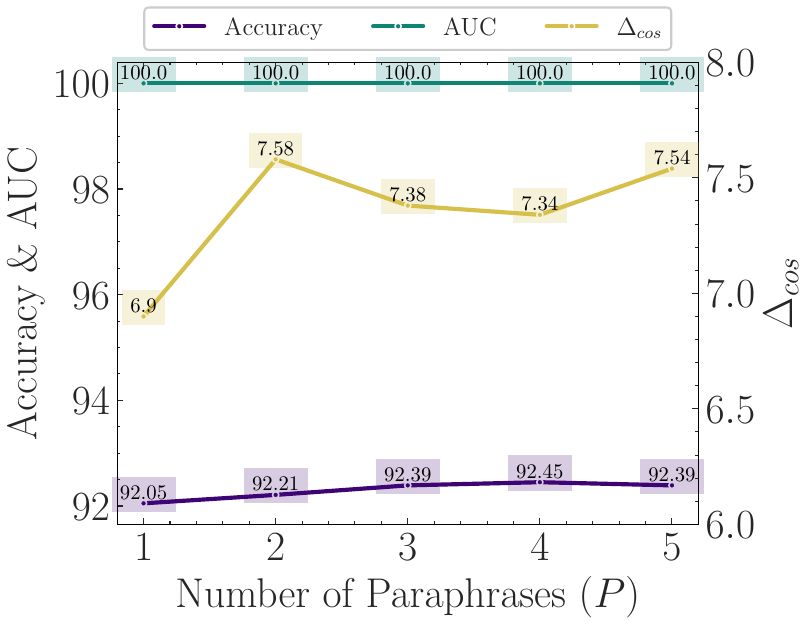}
    \caption{\agnews}
    \end{subfigure}
    \caption{\gpt paraphrase attack performance different number of paraphrases ($P$) for all the datasets.}
    \label{fig:diff-P-GPT-attack}
\end{figure*}

\begin{figure*}[h!]
    \centering
    \begin{subfigure}{0.24\textwidth}
    \centering\includegraphics[width=\linewidth,keepaspectratio]{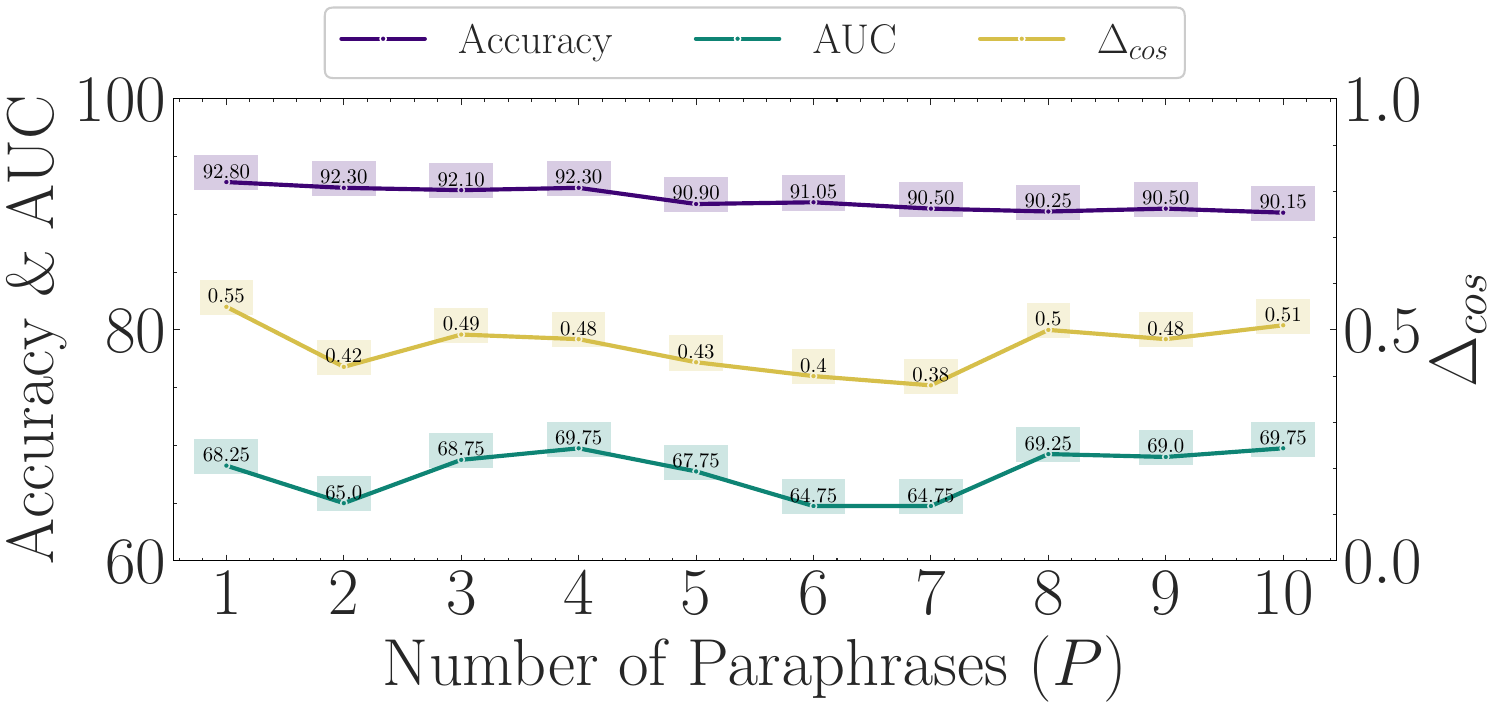}
    \caption{\enron}
    \end{subfigure}
    \begin{subfigure}{0.24\textwidth}
    \centering\includegraphics[width=\linewidth,keepaspectratio]{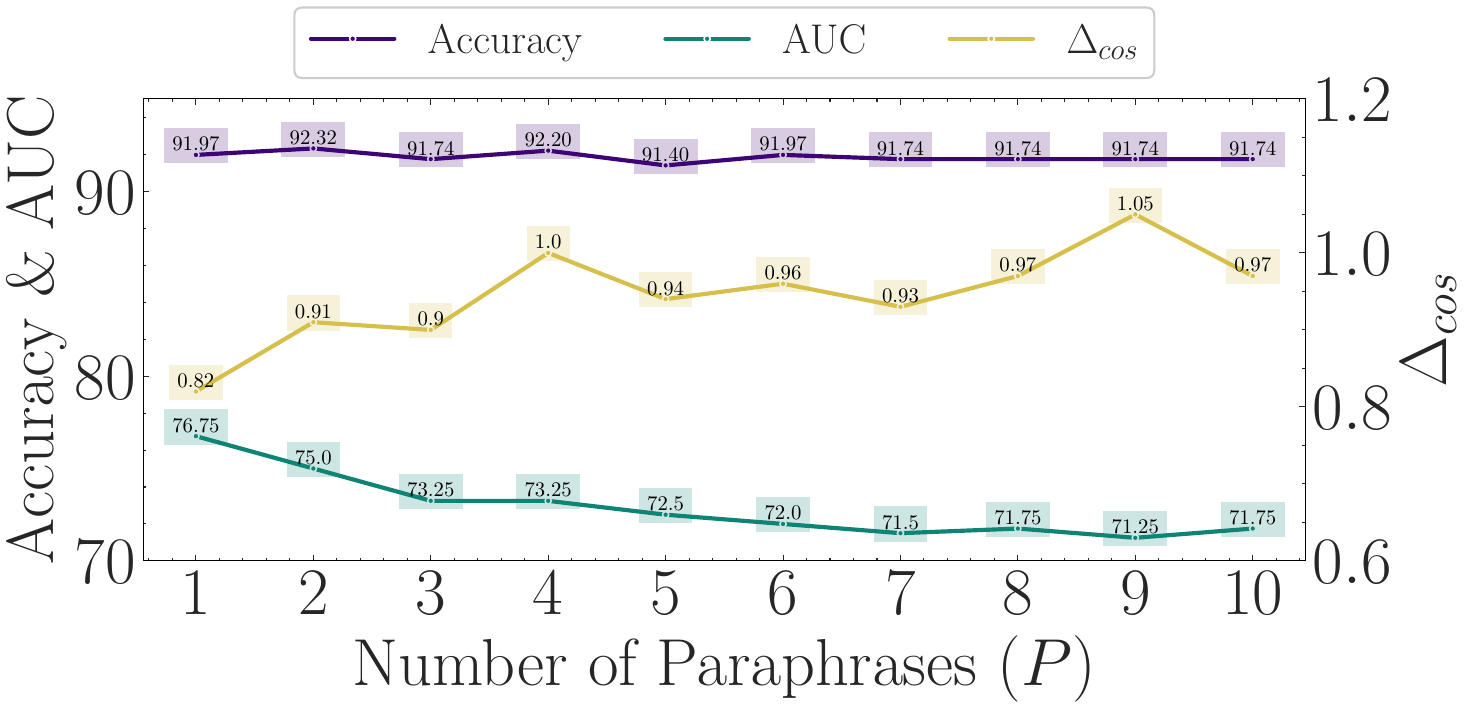}
    \caption{\sst}
    \end{subfigure}
    \begin{subfigure}{0.24\textwidth}
    \centering\includegraphics[width=\linewidth,keepaspectratio]{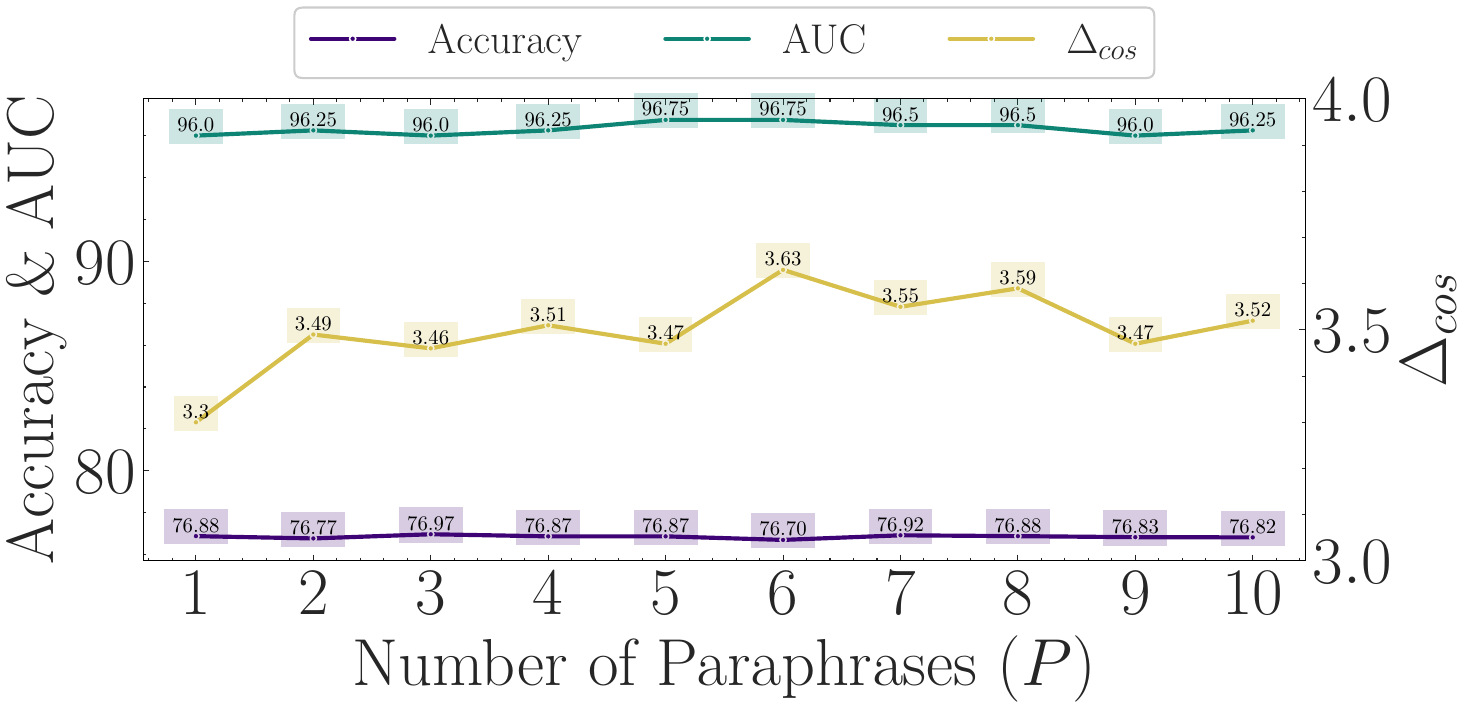}
    \caption{\mind}
    \end{subfigure}
    \begin{subfigure}{0.24\textwidth}
    \centering\includegraphics[width=\linewidth,keepaspectratio]{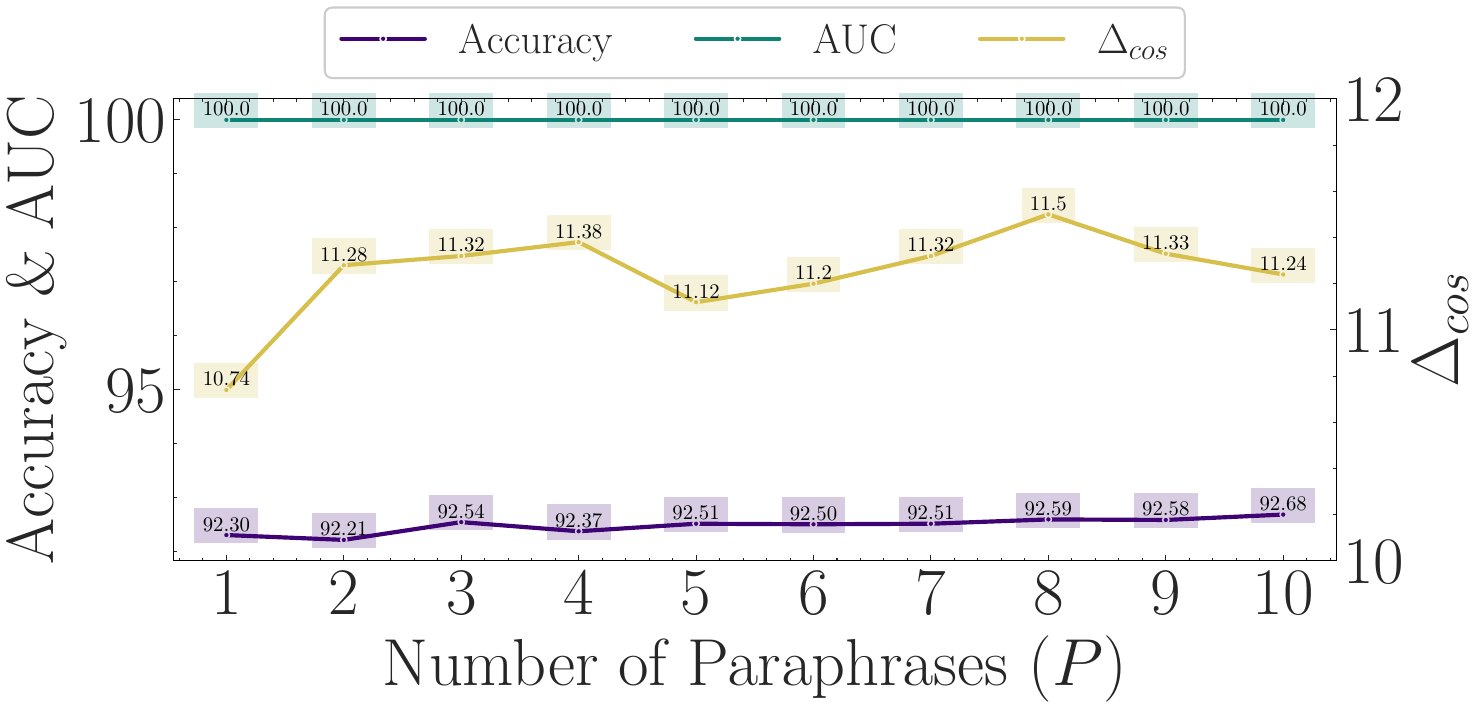}
    \caption{\agnews}
    \end{subfigure}
    \caption{DIPPER paraphrase attack performance different number of paraphrases ($P$) for all the datasets. 
    }
    \label{fig:diff-P-DIPPER-attack}
\end{figure*}

\begin{figure*}[h!]
    \centering
    \begin{subfigure}{0.235\textwidth}
    \centering\includegraphics[width=\linewidth,keepaspectratio]{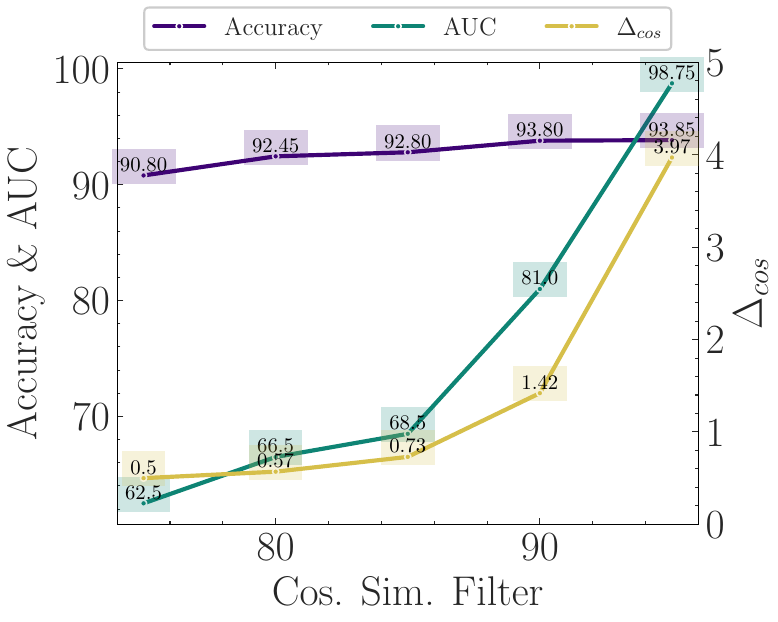}
    \caption{\enron}
    \end{subfigure}
    \begin{subfigure}{0.24\textwidth}
    \centering\includegraphics[width=\linewidth,keepaspectratio]{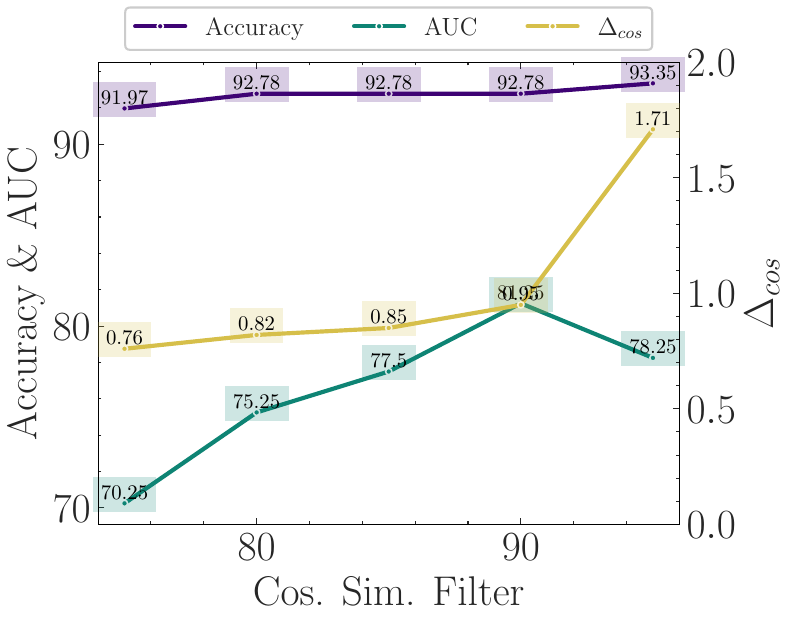}
    \caption{\sst}
    \end{subfigure}
    \begin{subfigure}{0.25\textwidth}
    \centering\includegraphics[width=\linewidth,keepaspectratio]{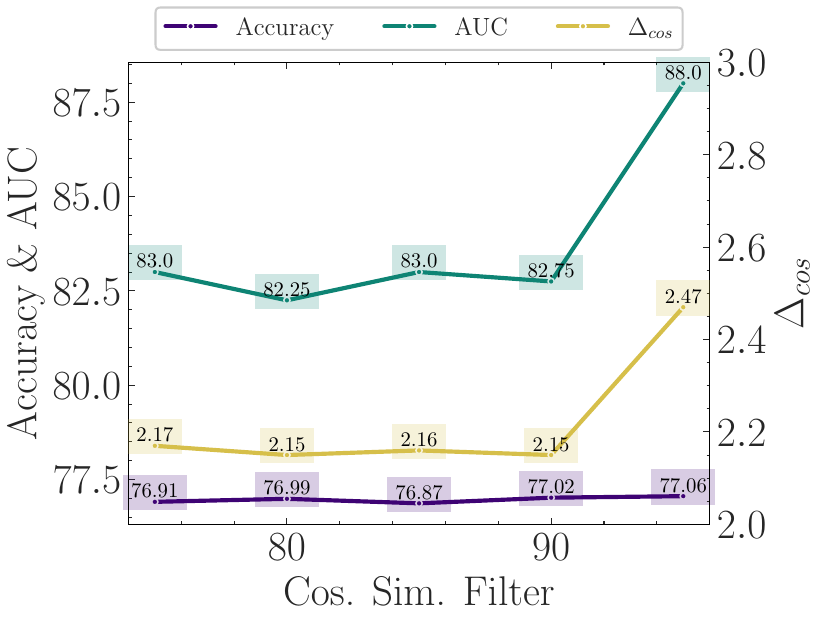}
    \caption{\mind}
    \end{subfigure}
    \begin{subfigure}{0.245\textwidth}
    \centering\includegraphics[width=\linewidth,keepaspectratio]{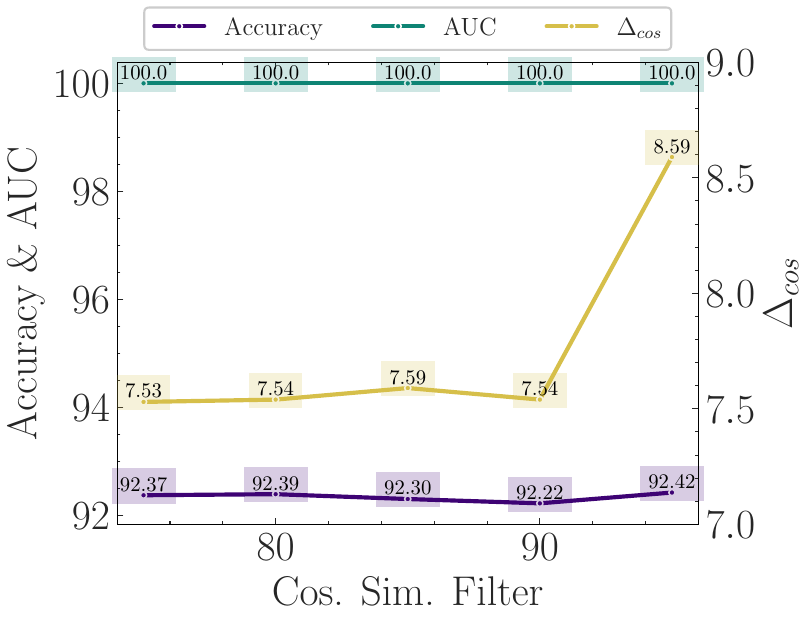}
    \caption{\agnews}
    \end{subfigure}
    \caption{\gpt paraphrase attack performance using different cosine similarity filters for all the datasets.}
    \label{fig:cos-filter}
\end{figure*}

\begin{figure*}[h!]
    \centering
    \begin{subfigure}{0.32\textwidth}
    \centering\includegraphics[width=\linewidth,keepaspectratio]{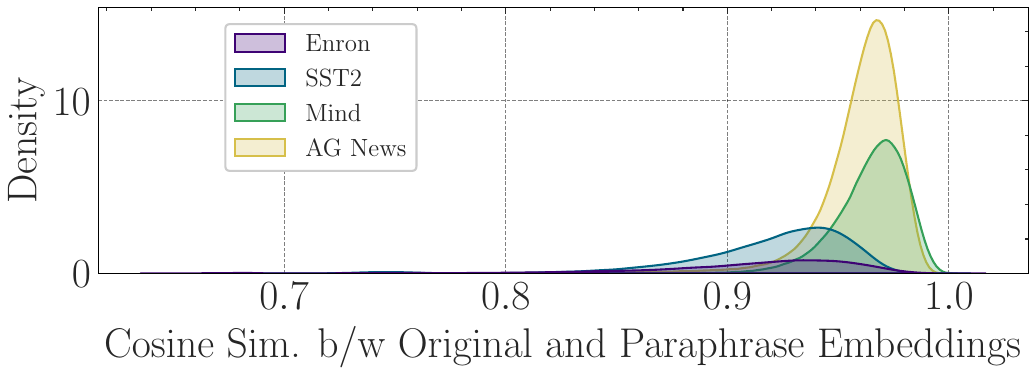}
    \caption{\gpt}
    \end{subfigure}
    \begin{subfigure}{0.33\textwidth}
    \centering\includegraphics[width=\linewidth,keepaspectratio]{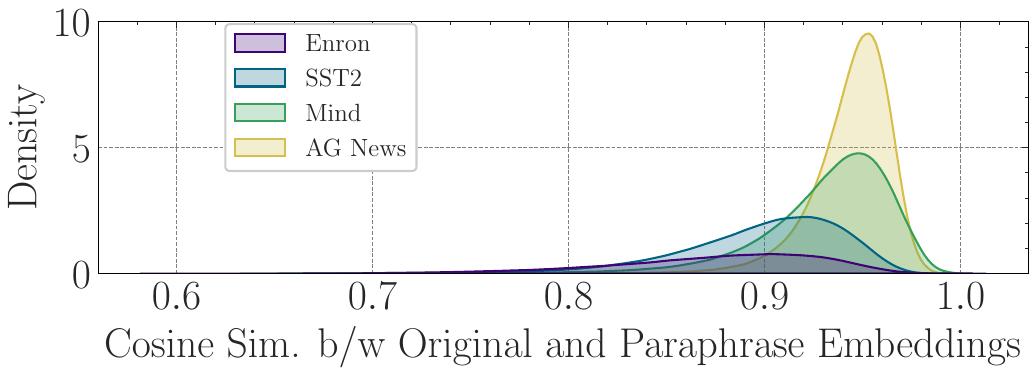}
    \caption{DIPPER}
    \end{subfigure}
    \begin{subfigure}{0.32\textwidth}
    \centering\includegraphics[width=\linewidth,keepaspectratio]{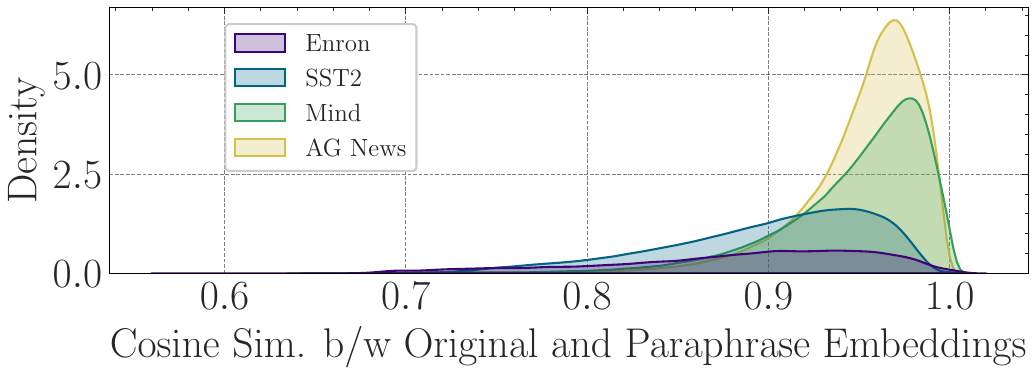}
    \caption{\nllb}
    \end{subfigure}
    \caption{Cosine similarity between original and paraphrased embeddings using different paraphrases (as denoted in the captions).}
    \label{fig:cos-sim-paras}
\end{figure*}

In Figures~\ref{fig:diff-P-GPT-attack}~and~\ref{fig:diff-P-DIPPER-attack}, we study the influence of the number of paraphrases ($P$) in \ourattack. We observe there are no significant changes in attack performance with $\uparrow P$. This shows lower number of paraphrases might suffice in the attack.

\subsection{Non-Watermark Case}
\label{app:attack-non-wm}
\begin{table}[h]
\centering
    \begin{minipage}{0.95\columnwidth}
    \resizebox{\textwidth}{!}{%
    \begin{tabular}{ccc@{\;\;}cc@{\;\;}c@{\;\;}}
    \toprule
    \multirow{2}{*}{Dataset} & \multirow{2}{*}{Paraphraser} & \multicolumn{2}{c}{Utility} & \multicolumn{2}{c}{Verifiability} \\
    \cmidrule(lr){3-4} \cmidrule(lr){5-6}
    {} & {} & ACC $\uparrow$ & F1 $\uparrow$ & {$\Delta_{cos} \downarrow$} & {AUC $\downarrow$} \\
    \toprule
    \multirow{3}{*}{\enron}
         {} & {\gpt} & 92.85 & 92.85 & 0.32 & 57.25 \\
         {} & {\dipper} & 92.00 & 91.99 & -0.18 & 42.25 \\
         {} & {\nllb} & 93.25 & 93.25 & 0.27 & 58.00 \\
    \midrule
    \multirow{3}{*}{\sst}
         {} & {\gpt}  & 92.43 & 92.43 & 0.65 & 70.75 \\
         {} & {\dipper} & 91.17 & 91.13 & 0.54 & 67.50 \\
         {} & {\nllb} & 92.20 & 92.18 & 0.22 & 59.50 \\
    \midrule
    \multirow{3}{*}{\mind}
        {} & {\gpt}  & 76.97 & 51.29 & 1.09 & 70.75 \\
        {} & {\dipper} & 76.98 & 50.68 & 1.23 & 76.25 \\
        {} & {\nllb} & 76.72 & 50.47 & 1.17 & 75.00 \\
    \midrule
    \multirow{3}{*}{\agnews}
        {} & {\gpt} & 92.22 & 92.21 & 1.87 & 88.50 \\
        {} & {\dipper} & 92.51 & 92.51 & 1.28 & 84.50 \\
        {} & {\nllb} & 92.61 & 92.59 & 1.35 & 85.25 \\
    \bottomrule
    \end{tabular}}
    \caption{Paraphrasing attack on a non-watermarked victim model. 
    }
    \label{table:attack-on-non-watermarked-model}
    \end{minipage}
\end{table}

It will be unknown to an attacker whether the model they are attempting to copy has watermarks. \reftab{table:attack-on-non-watermarked-model} demonstrates the suitability of \ourattack by causing minimal degradation in the utility and verifiability metrics.

\subsection{Impact of Attack Model Size}
\label{sec:attack-diff-model-size}
\begin{table}[h]
\begin{minipage}{0.95\columnwidth}
\resizebox{\textwidth}{!}{%
    \centering
    \begin{tabular}{ccccccc}
    \toprule
    \multirow{2}{*}{Dataset} & \multirow{2}{*}{Size} & \multicolumn{2}{c}{Utility} & \multicolumn{2}{c}{Verifiability} \\
    \cmidrule(lr){3-4} \cmidrule(lr){5-6}
    {} & {} & ACC $\uparrow$ & F1 $\uparrow$ & {$\Delta_{cos} \downarrow$} & {AUC $\downarrow$} \\
    \toprule
        \enron &  \multirow{4}{*}{Small} & 92.60 & 92.60 & 0.12 & 57.75 \\
        \sst & {}& 92.55 & 92.55 & 0.37 & 63.75 \\
        \mind & {} & 76.93 & 51.17 & 2.18 & 83.50 \\
        \agnews & {} & 92.42 & 92.40 & 8.61 & 100.00 \\
        \midrule
        \enron &  \multirow{4}{*}{Base} & 92.45 & 92.45 & 0.57 & 66.50 \\
        \sst & {}  & 92.78 & 92.77 & 0.82 & 75.25 \\
        \mind & {} & 76.99 & 51.48 & 2.15 & 82.25 \\
        \agnews & {} & 92.39 & 92.38 & 7.54 & 100.00 \\
        \midrule
        \enron &  \multirow{4}{*}{Large} & 92.70 & 92.70 & 0.39 & 61.50 \\
        \sst & {}  & 93.12 & 93.12 & 0.31 & 62.50 \\
        \mind & {} & 76.95 & 51.34 & 2.16 & 78.75 \\
        \agnews & {} & 92.51 & 92.49 & 7.29 & 100.00 \\
    \bottomrule
    \end{tabular}}
\caption{The impact of extracted model size on \ourattack performance. 
}
\label{table:attack-model-sizes}
\end{minipage}
\end{table}

We assess whether our attack's performance varies with the attacker model's size. We conducted experiments for \gpt~\ourattack using small, base, and large variants of the BERT \citep{devlin-etal-2019-bert} model to test this. The results, summarised in the \reftab{table:attack-model-sizes}, demonstrate that the attack consistently circumvents the watermark, regardless of the model size.

\subsection{Impact of Scaling Train Dataset}
\label{sec:scale-up-dataset}
\begin{table}[h]
    \begin{minipage}{0.95\columnwidth}
    \resizebox{\textwidth}{!}{%
    \begin{tabular}{cccccc}
    \toprule
    \multirow{2}{*}{Dataset} & \multirow{2}{*}{Type} & \multicolumn{2}{c}{Utility} & \multicolumn{2}{c}{Verifiability} \\
    \cmidrule(lr){3-4} \cmidrule(lr){5-6}
    {} & {} & ACC $\uparrow$ & F1 $\uparrow$ & {$\Delta_{cos} \downarrow$} & {AUC $\downarrow$} \\
    \toprule
    
    \multirow{3}{*}{\enron}
         {} & {\gpt} & 95.30 & 95.30 & 6.65 & 98.50 \\
         {} & {\dipper} & 95.30 & 95.30 & 8.47 & 99.50 \\
         {} & {\nllb} & 94.95 & 94.95 & 8.64 & 99.25 \\
    \midrule
    \multirow{3}{*}{\sst}
         {} & {\gpt} & 93.35 & 93.34 & 6.69 & 96.25 \\
         {} & {\dipper} & 92.66 & 92.65 & 8.73 & 99.50 \\
         {} & {\nllb} & 93.23 & 93.23 & 7.36 & 98.25 \\
    \midrule
    \multirow{3}{*}{\mind}
        {} & {\gpt} & 77.06 & 52.07 & 12.74 & 100.00 \\
        {} & {\dipper} & 77.23 & 55.46 & 15.58 & 100.00 \\
        {} & {\nllb} & 77.12 & 56.89 & 14.97 & 100.00 \\
    \midrule
    \multirow{3}{*}{\agnews}
        {} & {\gpt} & 93.11 & 93.10 & 19.68 & 100.00 \\
        {} & {\dipper} & 93.59 & 93.58 & 19.26 & 100.00 \\
        {} & {\nllb} & 93.39 & 93.39 & 18.97 & 100.00 \\

    \bottomrule
    \end{tabular}}
    \caption{The impact of scaling up the dataset with paraphrases instead of averaging the paraphrased embeddings in \ourattack. 
    }
    \label{fig:scale-up-dataset}
    \end{minipage}
\end{table}

A potential confound is that creating multiple paraphrases effectively increases the training data size during the imitation attack. To tease out this effect, we run another experiment where we scale up the training data size to match the size used in the paraphrasing experiment; results in \reftab{fig:scale-up-dataset}. Interestingly, we found that watermark detection performance goes up, showing that the success of paraphrasing in evading detection is not due to increased training data size.

\subsection{Filtering with Different Cosine Similarity}
\label{appendix:cos-sim-filter}

We observed that the Enron dataset contains derogatory or ambiguous text to which \gpt responds with a general disclaimer or refusal to answer. We implemented a filtering process to ensure that only relevant content is used. If no valid paraphrases are found after filtering, we revert to the default response. We also conducted an ablation study (more in \reffig{fig:cos-filter}) to determine the optimal cosine similarity threshold. We settle on 80\% for filtering providing a good tradeoff between quality and attack performance. This process effectively filters out texts with low paraphrase similarity, such as derogatory content or extremely vague and short texts like ``swill'' or ``free'' in the SST2 dataset.

\subsection{Quality of Paraphrases}
\label{app:para-quality}

Although the quality of paraphrases is not crucial for the attack, poor-quality paraphrases can result in utility loss. We did not conduct a human evaluation, as all the paraphrasers we use have already been evaluated for the same. Hence, we check the cosine similarity between original and paraphrased text embeddings.
We can note from \reffig{fig:cos-sim-paras} that most paraphrases are similar to the original text demonstrating good-quality paraphrases. Furthermore, with the implemented cosine similarity filter (as discussed in \refapp{appendix:cos-sim-filter}) we will remove low-quality paraphrases corresponding to left-side entries of the distribution plots.

\section{\ourdefence Analyses}
\label{sec:ablation-defence}
We perform detailed ablation studies for \ourdefence.

\subsection{Impact of Gaussian Noise}
\label{app:gauss-noise-defence}
\begin{table}[h!]
\centering
    \begin{minipage}{0.99\columnwidth}
    \resizebox{\textwidth}{!}{%
    \begin{tabular}{ccc@{\;\;}cc@{\;\;}c@{\;\;}}
    \toprule
    \multirow{2}{*}{Dataset} & \multirow{2}{*}{$\phi$} & \multicolumn{2}{c}{Utility} & \multicolumn{2}{c}{Verifiability} \\
    \cmidrule(lr){3-4} \cmidrule(lr){5-6}
    {} & {} & ACC $\uparrow$ & F1 $\uparrow$ & {$\Delta_{cos} \uparrow$} & {AUC $\uparrow$} \\
    \toprule
    \multirow{5}{*}{\enron}
        & 0.01 & 93.45 & 93.45 & 62.29 & 100.00 \\
        & 0.05 & 84.00 & 84.00 & 17.39 & 100.00 \\
        & 0.10 & 73.60 & 73.59 & 8.84 & 99.00 \\
        & 0.50 & 52.30 & 51.30 & 1.77 & 70.62 \\
        & 1.00 & 50.95 & 49.52 & 0.88 & 61.71 \\
    \midrule

    \multirow{5}{*}{\sst}
        & 0.01 & 91.40 & 91.39 & 64.78 & 100.00 \\
        & 0.05 & 84.29 & 84.26 & 18.09 & 100.00 \\
        & 0.10 & 73.74 & 73.65 & 9.17 & 99.55 \\
        & 0.50 & 53.67 & 49.87 & 1.79 & 69.03 \\
        & 1.00 & 51.72 & 45.76 & 0.86 & 59.32 \\
    \midrule

    \multirow{5}{*}{\mind}
        & 0.01 & 70.37 & 44.18 & 63.61 & 100.00 \\
        & 0.05 & 63.20 & 33.34 & 17.76 & 100.00 \\
        & 0.10 & 49.83 & 15.87 & 9.03 & 99.26 \\
        & 0.50 & 31.60 & 4.85 & 1.82 & 69.78 \\
        & 1.00 & 29.34 & 4.85 & 0.92 & 60.60 \\
    \midrule
    \multirow{5}{*}{\agnews}
        & 0.01 & 92.28 & 92.25 & 64.09 & 100.00 \\
        & 0.05 & 83.92 & 83.84 & 17.73 & 100.00 \\
        & 0.10 & 65.58 & 65.52 & 9.00 & 99.33 \\
        & 0.50 & 30.00 & 29.88 & 1.79 & 69.59 \\
        & 1.00 & 25.16 & 25.00 & 0.89 & 59.92 \\
    \bottomrule
    \end{tabular}}
    \end{minipage}
    \caption{Impact of different Gaussian noise ($\phi$) in \ourdefence for all the datasets.}
    \label{table:gauss-noise-WET}
\end{table}

We want to evaluate the effect of adding Gaussian noise to embeddings on watermark verification and downstream utility. Following \citep{morris-etal-2023-text,chen2024text}, we consider different noise levels ($\lambda$) and add noise as follows:

\begin{equation}
    \begin{aligned}
    \phi_{noisy}(x) = \text{Norm}(\phi(x) + \lambda \cdot \mathbf{\epsilon}), \mathbf{\epsilon} \sim \mathcal{N}(0, 1).
    \notag
    \end{aligned}
\end{equation}

How much perturbation can be handled by \ourdefence? From \reftab{table:gauss-noise-WET}, we can see that from $\phi=0.05$, we start seeing significant utility loss; however, we have a perfect AUC for this case for all the datasets. It demonstrates that \ourdefence has more capacity to handle perturbations and is more robust.

\subsection{Impact of Size of Verification Dataset}
\label{app:diff-verif-size-defence}
\begin{table}[h!]
\centering
    \begin{minipage}{0.85\columnwidth}
    \resizebox{\textwidth}{!}{%
    \begin{tabular}{cccc}
    \toprule
    \multirow{2}{*}{Dataset} & \multirow{2}{*}{$V$} & \multicolumn{2}{c}{Verifiability} \\
    \cmidrule(lr){3-4}
    {} & {} & {$\Delta_{cos} \uparrow$} & {AUC $\uparrow$} \\
    \toprule
    \multirow{6}{*}{\enron}
        & 1 & 90.00 & 100.00 \\
        & 5 & 88.26 & 100.00 \\
        & 20 & 89.90 & 100.00 \\
        & 100 & 89.14 & 100.00 \\
        & 500 & 89.67 & 100.00 \\
        & 1000 & 89.34 & 100.00 \\
    \midrule

    \multirow{5}{*}{\sst}
        & 1 & 90.29 & 100.00 \\
        & 5 & 93.75 & 100.00 \\
        & 20 & 93.67 & 100.00 \\
        & 100 & 93.77 & 100.00 \\
        & 436 & 93.89 & 100.00 \\
    \midrule

    \multirow{6}{*}{\mind}
        & 1 & 92.28 & 100.00 \\
        & 5 & 89.90 & 100.00 \\
        & 20 & 90.74 & 100.00 \\
        & 100 & 90.54 & 100.00 \\
        & 500 & 90.97 & 100.00 \\
        & 1000 & 90.87 & 100.00 \\
    \midrule

    \multirow{6}{*}{\agnews}
        & 1 & 96.13 & 100.00 \\
        & 5 & 93.62 & 100.00 \\
        & 20 & 92.63 & 100.00 \\
        & 100 & 91.77 & 100.00 \\
        & 500 & 92.10 & 100.00 \\
        & 1000 & 92.00 & 100.00 \\
    \bottomrule
    \end{tabular}}
    \end{minipage}
    \caption{Impact of different size of verification dataset sizes ($V$) in \ourdefence verifiability for all the datasets. Note: \sst has only 872 test samples (see \reftab{table:dataset-statistics}).}
    \label{tab:diff-v}
\end{table}

In this, we investigate the number of samples ($V$) we need in the verification dataset. From \reftab{tab:diff-v}, we can observe that \ourdefence's verification technique is not dependent on the number of verification samples, even just a single verification sample might suffice. This is another advantage of our technique, do note in our experiment we use $v=250$ unless specified otherwise.

\subsection{Different Transformation Matrices Construction}
\label{app:diff-matrices}
\begin{table}[h]
\centering
    \begin{minipage}{0.99\columnwidth}
    \resizebox{\textwidth}{!}{%
    \begin{tabular}{ccc@{\;\;}cc@{\;\;}c@{\;\;}}
    \toprule
    \multirow{2}{*}{Dataset} & \multirow{2}{*}{Type} & \multicolumn{2}{c}{Utility} & \multicolumn{2}{c}{Verifiability} \\
    \cmidrule(lr){3-4} \cmidrule(lr){5-6}
    {} & {} & ACC $\uparrow$ & F1 $\uparrow$ & {$\Delta_{cos} \uparrow$} & {AUC $\uparrow$} \\
    \toprule
    \multirow{6}{*}{\enron}
        & Circulant & 94.75 & 94.75 & 89.22 & 100.00 \\
        \cmidrule{2-6}
        & New Wts. Circulant & 94.60 & 94.60 & 21.60 & 99.96 \\
        & Random & 94.30 & 94.30 & 22.95 & 99.96 \\
        & Eq. Wts. Circulant & 94.40 & 94.40 & 92.81 & 100.00 \\
        & Seq. Pos. Circulant  & 93.40 & 93.40 & 69.91 & 100.00 \\
        & Seq. Pos. and Eq. Wts. Circulant & 92.45 & 92.45 & -0.23 & 47.69 \\
    \midrule

    \multirow{6}{*}{\sst}
        & Circulant & 93.35 & 93.34 & 93.70 & 100.00 \\
       \cmidrule{2-6}
        & New Wts. Circulant & 93.00 & 93.00 & 23.60 & 99.99 \\
        & Random & 92.55 & 92.54 & 25.02 & 100.00 \\
        & Eq. Wts. Circulant & 92.78 & 92.77 & 96.13 & 100.00 \\
        & Seq. Pos. Circulant & 91.97 & 91.97 & 74.31 & 100.00 \\
        & Seq. Pos. and Eq. Wts. Circulant & 90.60 & 90.59 & 0.72 & 53.88 \\
    \midrule

    \multirow{6}{*}{\mind}
        & Circulant  & 77.21 & 51.36 & 91.12 & 100.00 \\
        \cmidrule{2-6}
        & New Wts. Circulant & 76.97 & 50.83 & 23.56 & 100.00 \\
        & Random & 77.00 & 50.94 & 24.82 & 100.00 \\
        & Eq. Wts. Circulant & 77.04 & 51.03 & 95.34 & 100.00 \\
        & Seq. Pos. Circulant & 76.61 & 50.06 & 71.93 & 100.00 \\
        & Seq. Pos. and Eq. Wts. Circulant & 75.21 & 47.22 & 0.02 & 50.28 \\
        
    \midrule

    \multirow{6}{*}{\agnews}
        & Circulant & 93.03 & 93.02 & 92.05 & 100.00 \\
        \cmidrule{2-6}
        & New Wts. Circulant & 93.20 & 93.19 & 26.60 & 100.00 \\
        & Random & 92.95 & 92.94 & 27.55 & 100.00 \\
        & Eq. Wts. Circulant & 93.07 & 93.06 & 96.47 & 100.00 \\
        & Seq. Pos. Circulant & 92.41 & 92.40 & 73.79 & 100.00 \\
        & Seq. Pos. and Eq. Wts. Circulant & 91.89 & 91.88 & -0.27 & 50.37 \\

    \bottomrule
    \end{tabular}}
    \end{minipage}
    \caption{\ourdefence performance using different variation of transformation matrix $\rmT$ as defined in the \refsec{app:diff-matrices}.}
    \label{table:diff-matrices}
\end{table}

We investigate different alterations to our construction of the transformation matrix $\rmT$, discussed in \refsec{sec:matrix-construction}. 

\paragraph{New Wts. Circulant Matrix.} In this, we construct new weights for each row in the circulant matrix. However, with this, we also lose the full-rank and well-conditioned properties.

\paragraph{Random Matrix.} This is a pure random generation process where we randomly pick $k$ non-zero positions and assign random values to them, for each row. 

\paragraph{Eq. Wts. Circulant.} We set the $1/k$ as the value to non-zero positions in the row. %

\paragraph{Seq. Pos. Circulant.} We pick the first $k$ dimensions in the row as the non-zero positions.

\paragraph{Seq. Pos. and Eq. Wts. Circulant.} This is the combination of the previous two matrix constructions.

\paragraph{Discussion.}
We present the performance of such matrices in \reftab{table:diff-matrices}. We see that equal weights and sequential position-based matrices have strong verifiability. However, such matrix constructions are not stealthy with equal weights or sequential positions. At the same time, the matrix combining the above methods is much worse in terms of verifiability. The other two constructions of new row weights every time in circulant matrix and pure random matrix construction have low $\Delta_{cos}$ metric even though it has perfect AUC. The reason is such matrices are not full-rank and are not well-conditions leading to poorer reverse transformation.

\subsection{Impact of Attack Model Size}
\label{app:diff-model-size-defence}
\begin{table}[h]
\begin{minipage}{0.95\columnwidth}
\resizebox{\textwidth}{!}{%
    \centering
    \begin{tabular}{cccccccccc}
    \toprule
    \multirow{3}{*}{Dataset} & \multirow{3}{*}{Size} & \multicolumn{2}{c}{Utility} & \multicolumn{2}{c}{Verifiability} \\
    \cmidrule(lr){3-4} \cmidrule(lr){5-6}
    {} & {} & ACC $\uparrow$ & F1 $\uparrow$ & {$\Delta_{cos} \uparrow$} & {AUC $\uparrow$} \\
    \toprule
        \enron &  \multirow{4}{*}{Small} & 94.55 & 94.55 & 88.74 & 100.00 \\
        \sst & {} & 93.23 & 93.23 & 93.12 & 100.00 \\
        \mind & {} & 77.15 & 51.00 & 90.25 & 100.00 \\
        \agnews & {} & 92.92 & 92.91 & 91.30 & 100.00 \\
        \midrule
        \enron &  \multirow{4}{*}{Base} & 94.75 & 94.75 & 89.22 & 100.00 \\
        \sst & {} & 93.35 & 93.34 & 93.70 & 100.00 \\
        \mind & {} & 77.21 & 51.36 & 91.12 & 100.00 \\
        \agnews & {} & 93.03 & 93.02 & 92.05 & 100.00 \\
        \midrule
        \enron &  \multirow{4}{*}{Large} & 94.40 & 94.40 & 88.32 & 100.00 \\
        \sst & {} & 93.00 & 93.00 & 93.60 & 100.00 \\
        \mind & {} & 76.95 & 50.77 & 90.94 & 100.00 \\
        \agnews & {} & 93.29 & 93.28 & 92.43 & 100.00 \\
    \bottomrule
    \end{tabular}}
\caption{The impact of extracted model size on \ourdefence performance.}
\label{table:WET-attack-model-sizes}
\end{minipage}
\end{table}

We assess whether our defence's performance varies
with the attacker model's size. We conducted experiments for \ourdefence using small, base, and large variants of the BERT model \citep{devlin-etal-2019-bert}. The results, summarised in the \reftab{table:WET-attack-model-sizes}, demonstrate that the defence works effectively with similar utility and verifiability.

\begin{figure*}[ht!]
    \centering
    \begin{subfigure}{0.32\textwidth}
        \includegraphics[width=\textwidth,keepaspectratio]{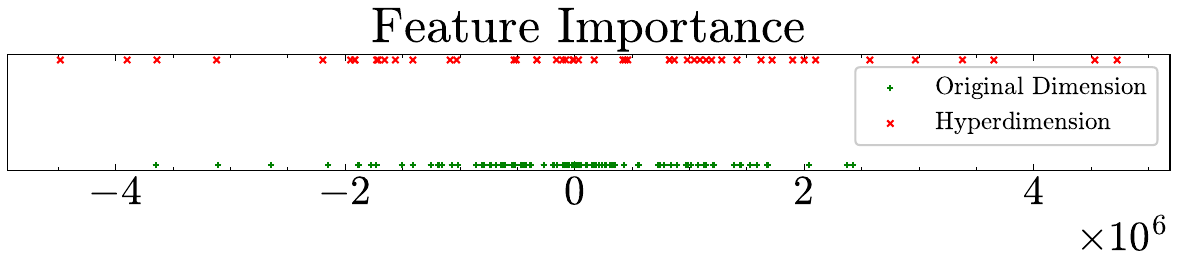}
        \caption{$k=2$}
    \end{subfigure}
    \begin{subfigure}{0.32\textwidth}
        \includegraphics[width=\textwidth,keepaspectratio]{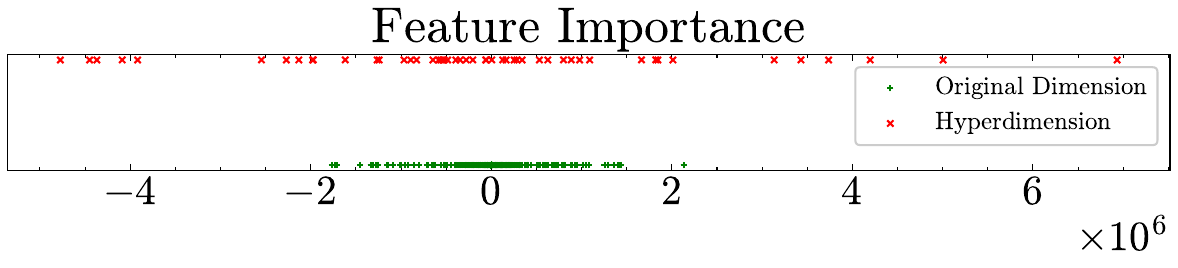}
        \caption{$k=5$}
    \end{subfigure}
    \begin{subfigure}{0.32\textwidth}
        \includegraphics[width=\textwidth,keepaspectratio]{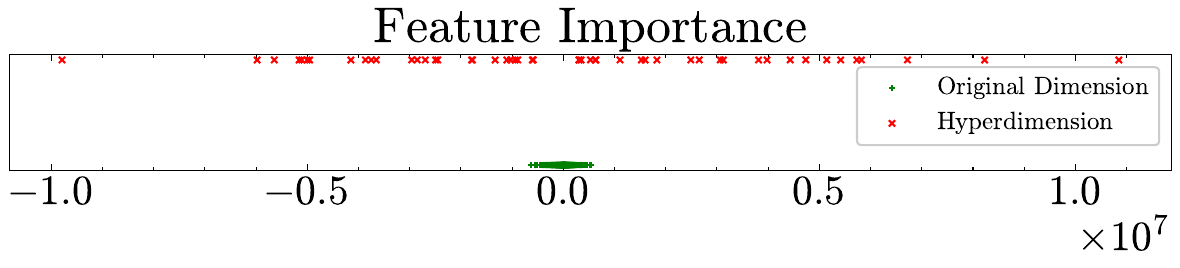}
        \caption{$k=50$}
    \end{subfigure}
    \begin{subfigure}{0.32\textwidth}
        \includegraphics[width=\textwidth,keepaspectratio]{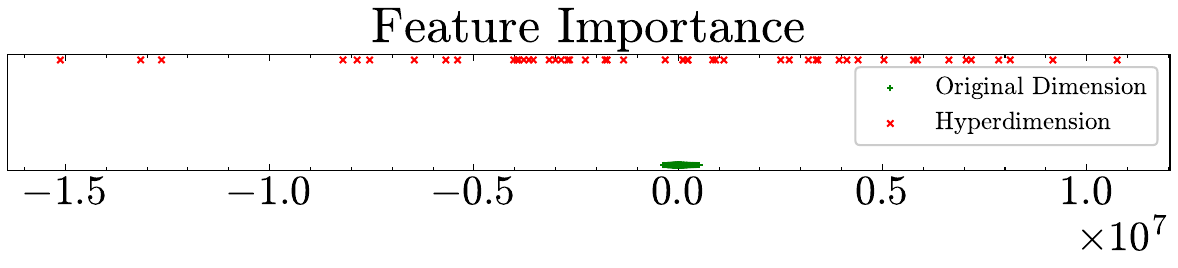}
        \caption{$k=100$}
    \end{subfigure}
    \begin{subfigure}{0.32\textwidth}
        \includegraphics[width=\textwidth,keepaspectratio]{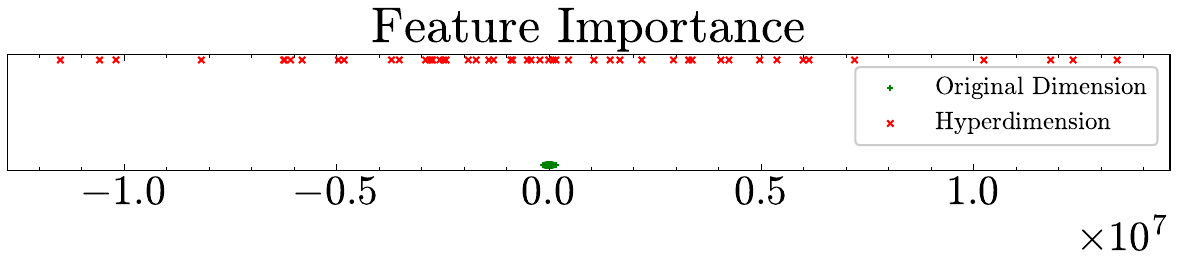}
        \caption{$k=500$}
    \end{subfigure}
    \begin{subfigure}{0.32\textwidth}
        \includegraphics[width=\textwidth,keepaspectratio]{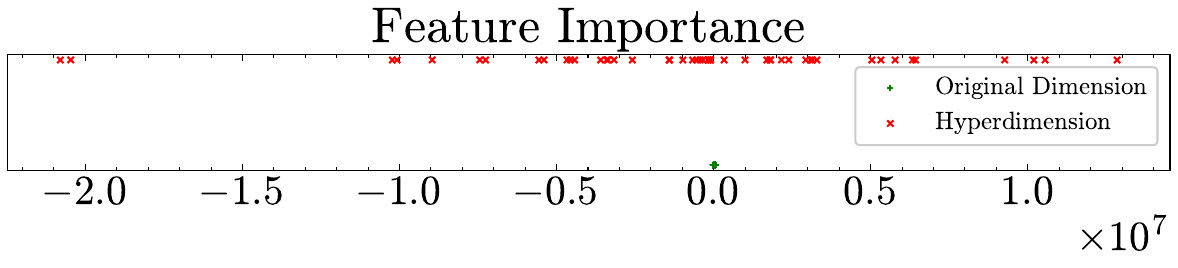}
        \caption{$k=1536$}
    \end{subfigure}
    \caption{Visualisation plots for feature importance of watermarked embedding dimensions in \sst.}
    \label{fig:feature-importance}
\end{figure*}

\begin{figure*}[ht!]
    \begin{subfigure}{0.32\textwidth}
        \includegraphics[width=\textwidth,keepaspectratio]{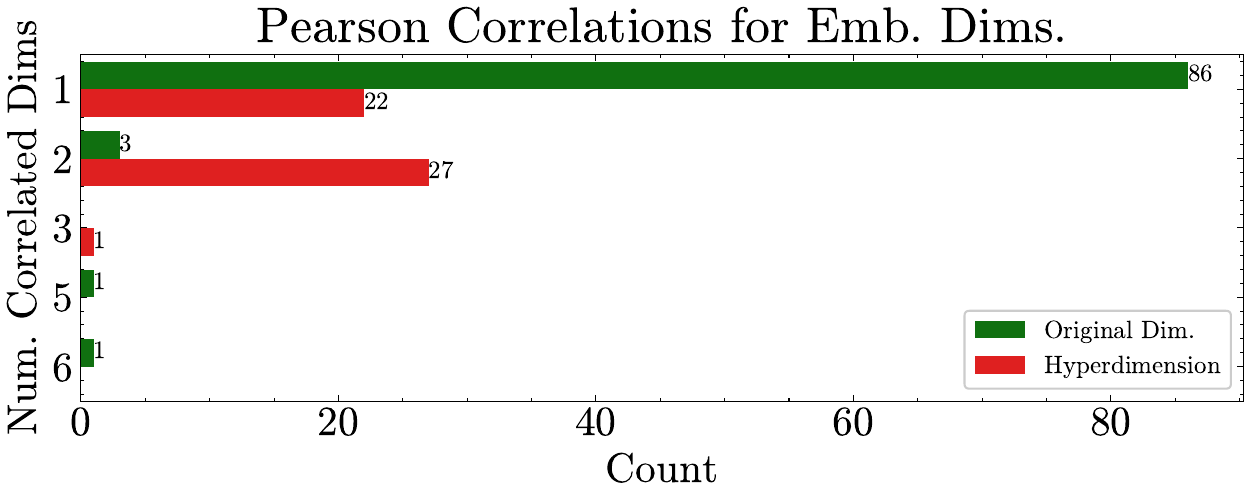}
        \caption{$k=2$}
    \end{subfigure}
    \begin{subfigure}{0.32\textwidth}
        \includegraphics[width=\textwidth,keepaspectratio]{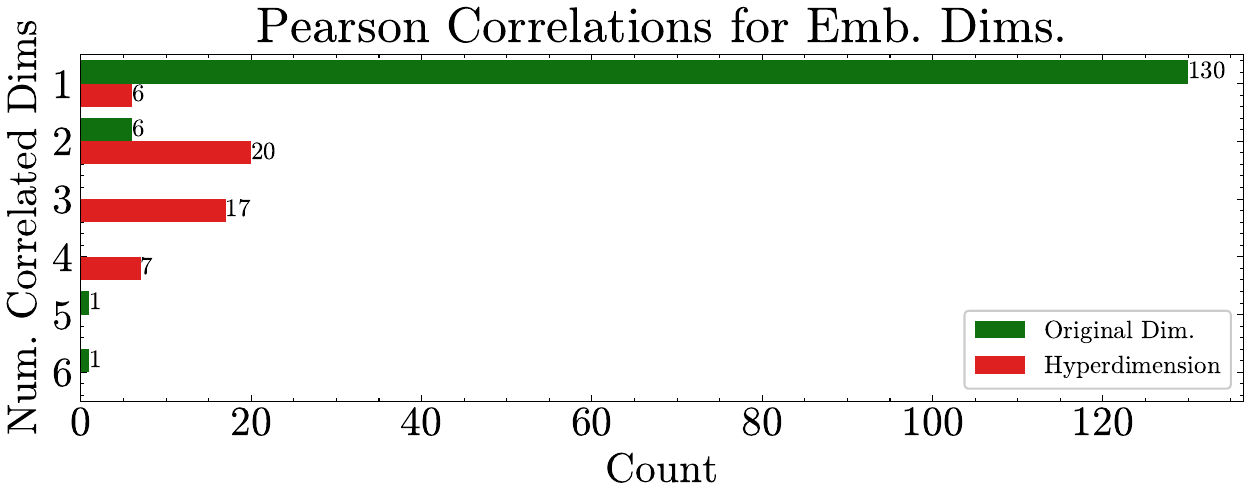}
        \caption{$k=5$}
    \end{subfigure}
    \begin{subfigure}{0.32\textwidth}
        \includegraphics[width=\textwidth,keepaspectratio]{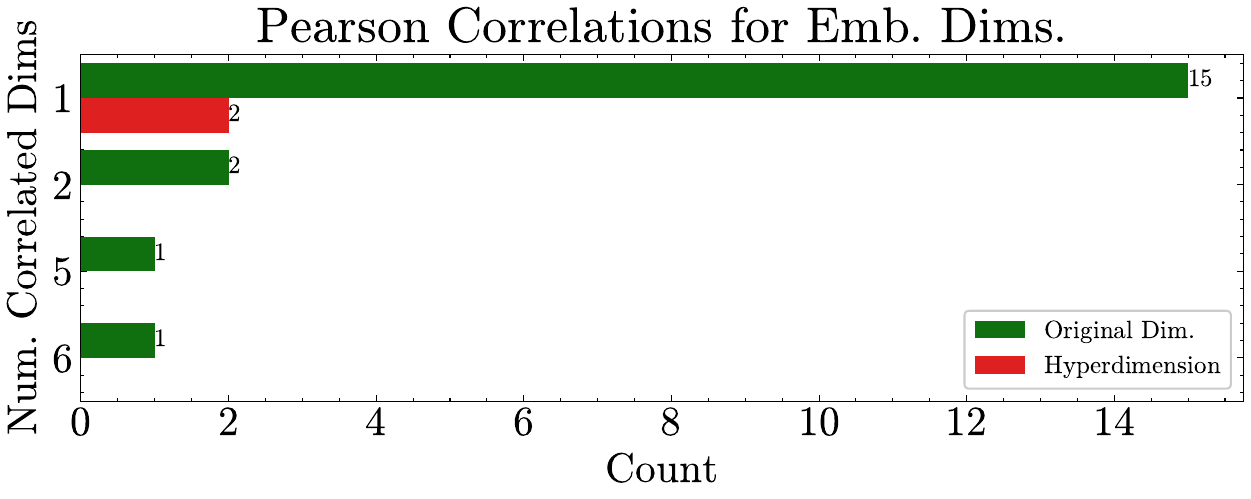}
        \caption{$k=50$}
    \end{subfigure}
    \begin{subfigure}{0.32\textwidth}
        \includegraphics[width=\textwidth,keepaspectratio]{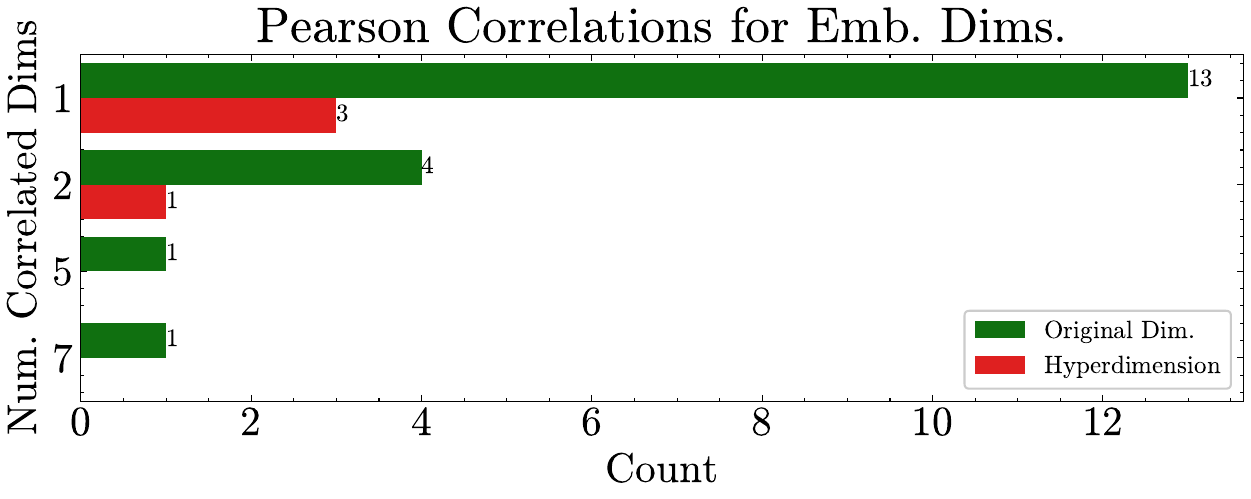}
        \caption{$k=100$}
    \end{subfigure}
    \hspace{.3em}
    \begin{subfigure}{0.32\textwidth}
        \includegraphics[width=\textwidth,keepaspectratio]{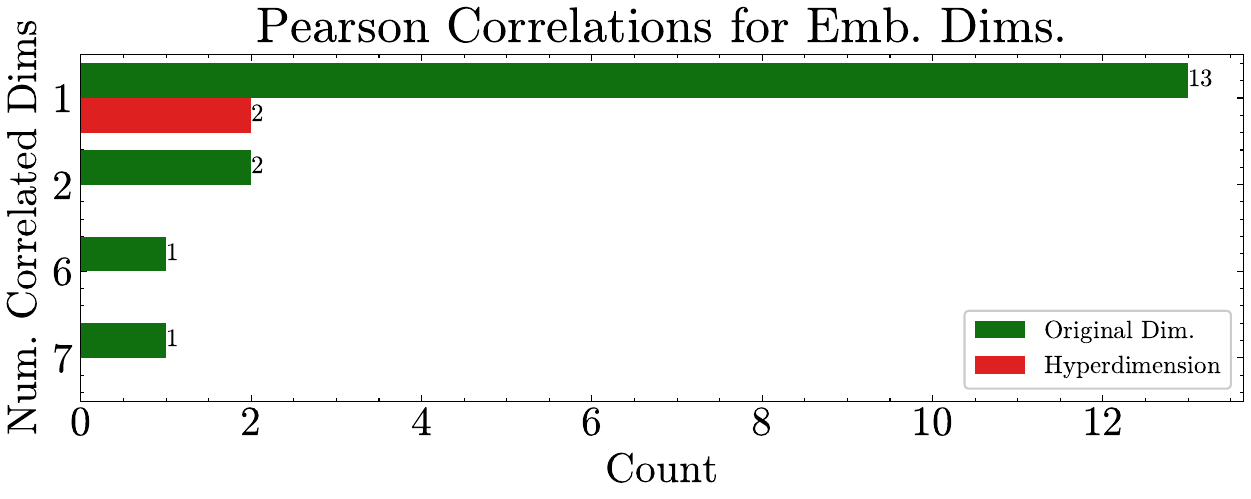}
        \caption{$k=500$}
    \end{subfigure}
    \hspace{.3em}
    \begin{subfigure}{0.32\textwidth}
        \includegraphics[width=\textwidth,keepaspectratio]{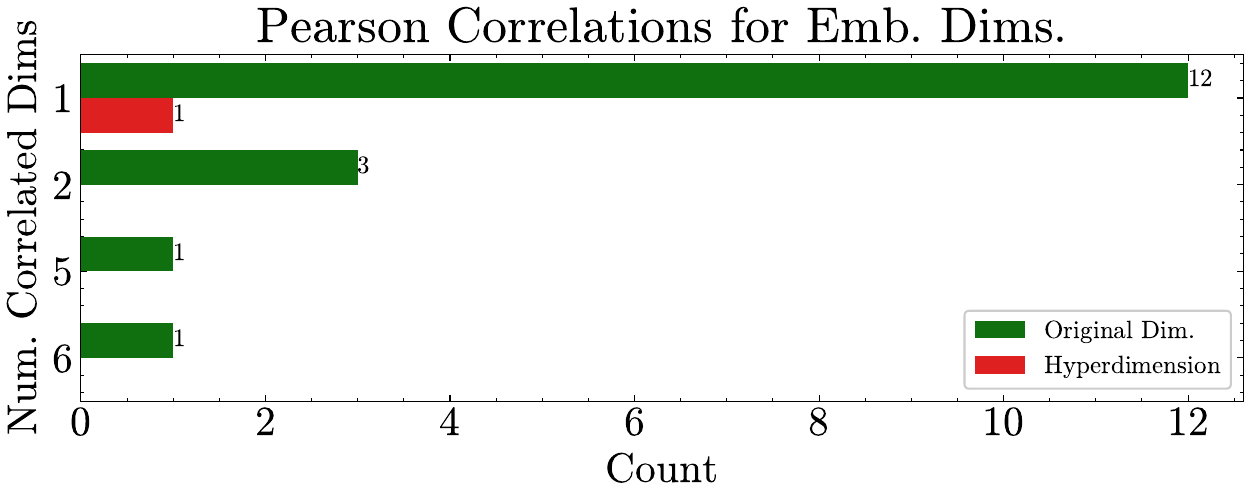}
        \caption{$k=1536$}
    \end{subfigure}
    \caption{Visualisation plots for feature correlations of watermarked embedding dimensions in \sst.}
    \label{fig:feature-correlations}
\end{figure*}

\subsection{Hyperdimension Obfuscation}
\label{sec:hyp-obfuscation}
In this section, we focus on the case where we add extra dimensions (\aka hyperdimensions; $w=50$) among the original embeddings. The positions of these hyperdimensions are randomly decided, and the value is a linear transformation of some $k$ existing original dimensions (similar to \refalg{algo:matrix-generation} but used as additional dimensions). For verification, we use the same ideas as in \ourdefence, with the only difference being that we work only on these obfuscated hyperdimensions. The utility and verifiability were comparable to \ourdefence. To evaluate the stealthiness of these hyperdimensions, we investigated feature correlation and feature importance techniques between hyperdimension and original dimensions. Properly mixed feature importance weights illustrate that hyperdimensions are indistinguishable. Similarly, uncorrelated hyperdimensions are appreciated, or else they are redundant. Note that these stealthiness techniques are not applicable for \ourdefence as we discard the original embedding dimensions.

\subsubsection{Feature Importance}

We train a linear regression (since we are working with linear transformations) with all the watermarked embeddings (original and hyperdimensions) for the downstream task. We use the weights of the linear regression as the feature importance weights. In \reffig{fig:feature-importance}, we represent these plots for different values of $k$. From this, we can conclude that we need $k<5$, as for higher values, hyperdimensions are discernible from the original embedding dimensions. For higher values of $k$, we have hyperdimensions that have more feature importance, which is logical considering linear combinations used in hyperdimension will represent the whole embedding with a higher value of $k$.

\subsubsection{Feature Correlations}

In this analysis, we use Pearson's coefficient \citep{sedgwick2012pearson} with a threshold of 0.4. The plots in \reffig{fig:feature-correlations} indicate that a $k$ value between 5 and 50 is required. However, this range conflicts with the values necessary ($k<5$) to bypass feature importance evaluation. Consequently, these plots (Figures~\ref{fig:feature-importance}~and~\ref{fig:feature-correlations}) lead us to conclude that hyperdimension obfuscation will not work as they are easily detectable.

\begin{table}[t]
\centering
    \begin{minipage}{0.99\columnwidth}
    \resizebox{\linewidth}{!}{%
    \begin{tabular}{cccccccccc}
    \toprule
    \multirow{3}{*}{Dataset} & \multirow{3}{*}{$k$} & \multicolumn{2}{c}{Utility} & \multicolumn{2}{c}{Verifiability} \\
    \cmidrule(lr){3-4} \cmidrule(lr){5-6}
    {} & {} & ACC $\uparrow$ & F1 $\uparrow$ & {$\Delta_{cos} \uparrow$} & {AUC $\uparrow$} \\
    \toprule
    \multirow{9}{*}{\enron}
        & 1 & 94.75 & 94.75 & 89.13 & 100.00 \\
        & 2 & 94.75 & 94.75 & 82.92 & 100.00 \\
        & 5 & 94.80 & 94.80 & 87.64 & 100.00 \\
        & 25 & 94.75 & 94.75 & 89.22 & 100.00 \\
        & 50 & 94.40 & 94.40 & 90.86 & 100.00 \\
        & 100 & 94.35 & 94.35 & 82.84 & 100.00 \\
        & 500 & 92.85 & 92.85 & 81.70 & 100.00 \\
        & 1000 & 92.15 & 92.15 & 82.24 & 100.00 \\
        & 1536 & 91.70 & 91.70 & 85.50 & 100.00 \\
    \midrule

    \multirow{9}{*}{\sst}
        & 1 & 93.23 & 93.23 & 91.65 & 100.00 \\
        & 2 & 92.66 & 92.66 & 87.75 & 100.00 \\
        & 5 & 93.00 & 93.00 & 91.59 & 100.00 \\
        & 25 & 93.35 & 93.34 & 93.70 & 100.00 \\
        & 50 & 93.35 & 93.34 & 94.45 & 100.00 \\
        & 100 & 92.89 & 92.89 & 87.39 & 100.00 \\
        & 500 & 92.32 & 92.31 & 85.81 & 100.00 \\
        & 1000 & 92.55 & 92.54 & 86.58 & 100.00 \\
        & 1536 & 91.63 & 91.62 & 89.38 & 100.00 \\
    \midrule

    \multirow{9}{*}{\mind}
        & 1 & 77.25 & 51.40 & 91.62 & 100.00 \\
        & 2 & 77.10 & 51.19 & 85.70 & 100.00 \\
        & 5 & 77.16 & 51.05 & 89.19 & 100.00 \\
        & 25 & 77.21 & 51.36 & 91.12 & 100.00 \\
        & 50 & 76.95 & 50.71 & 92.41 & 100.00 \\
        & 100 & 76.88 & 50.72 & 85.06 & 100.00 \\
        & 500 & 76.61 & 49.85 & 85.42 & 100.00 \\
        & 1000 & 75.67 & 48.24 & 84.74 & 100.00 \\
        & 1536 & 74.28 & 42.96 & 85.46 & 100.00 \\
    \midrule

    \multirow{9}{*}{\agnews}
        & 1 & 93.45 & 93.44 & 92.85 & 100.00 \\
        & 2 & 93.46 & 93.46 & 86.68 & 100.00 \\
        & 5 & 93.22 & 93.22 & 90.59 & 100.00 \\
        & 25 & 93.03 & 93.02 & 92.05 & 100.00 \\
        & 50 & 93.22 & 93.22 & 93.30 & 100.00 \\
        & 100 & 93.00 & 93.00 & 86.90 & 100.00 \\
        & 500 & 92.62 & 92.61 & 86.28 & 100.00 \\
        & 1000 & 92.18 & 92.17 & 86.22 & 100.00 \\
        & 1536 & 91.59 & 91.58 & 86.66 & 100.00 \\

    \bottomrule
    \end{tabular}}
    \end{minipage}
    \caption{Different $k$ for $h=1536$ results. Expanding on \refsec{sec:diff-k-defence}, we provide detailed results here for completeness.}
    \label{table:diff-k-defence}
\end{table}

\begin{table}[h]
\centering
    \begin{minipage}{0.99\columnwidth}
    \resizebox{\linewidth}{!}{%
    \begin{tabular}{ccccccccccc}
    \toprule
    \multirow{2}{*}{Dataset} & \multirow{2}{*}{$w$} & \multicolumn{2}{c}{Utility} & \multicolumn{2}{c}{Verifiability} \\
    \cmidrule(lr){3-4} \cmidrule(lr){5-6}
    {} & {} & ACC $\uparrow$ & F1 $\uparrow$ & {$\Delta_{cos} \uparrow$} & {AUC $\uparrow$} \\
    \toprule
    \multirow{5}{*}{\enron}
        & 50 & 88.05 & 88.04 & 10.58 & 100.00 \\
        & 500 & 93.10 & 93.10 & 53.34 & 100.00 \\
        & 1000 & 94.15 & 94.15 & 74.95 & 100.00 \\
        & 1536 & 94.75 & 94.75 & 89.22 & 100.00 \\
        & 3000 & 94.75 & 94.75 & 90.49 & 100.00 \\
    \midrule

    \multirow{5}{*}{\sst}
        & 50 & 84.40 & 84.34 & 10.20 & 100.00 \\
        & 500 & 93.23 & 93.23 & 54.07 & 100.00 \\
        & 1000 & 93.58 & 93.57 & 77.07 & 100.00 \\
        & 1536 & 93.35 & 93.34 & 93.70 & 100.00 \\
        & 3000 & 92.55 & 92.54 & 94.47 & 100.00 \\
    \midrule

    \multirow{5}{*}{\mind}
        & 50 & 67.53 & 37.08 & 11.23 & 100.00 \\
        & 500 & 76.30 & 49.83 & 52.96 & 100.00 \\
        & 1000 & 76.78 & 50.55 & 75.74 & 100.00 \\
        & 1536 & 77.21 & 51.36 & 91.12 & 100.00 \\
        & 3000 & 76.96 & 50.72 & 93.85 & 100.00 \\
    \midrule

    \multirow{5}{*}{\agnews}
        & 50 & 85.07 & 85.04 & 12.44 & 100.00 \\
        & 500 & 92.46 & 92.45 & 53.55 & 100.00 \\
        & 1000 & 92.92 & 92.91 & 76.72 & 100.00 \\
        & 1536 & 93.03 & 93.02 & 92.05 & 100.00 \\
        & 3000 & 93.05 & 93.05 & 94.19 & 100.00 \\

    \bottomrule
    \end{tabular}}
    \end{minipage}
    \caption{Different $w$ for $k=25$ results. Expanding on \refsec{sec:diff-h-defence}, we provide detailed results here for completeness.}
    \label{table:diff-h-defence}
\end{table}

\end{document}